\documentclass[pra,aps,floatfix,10pt,twocolumn,nofootinbib,superscriptaddress]{revtex4-2} 
\usepackage[T1]{fontenc}
\usepackage{lmodern}
\usepackage[utf8]{inputenc}
\usepackage{microtype}
\usepackage{array}
\usepackage{graphicx}
\usepackage{dcolumn}
\usepackage{bm}
\usepackage{amsmath}
\usepackage{amsfonts}
\usepackage{amssymb}
\usepackage{amstext}   
\usepackage{amsthm}
\usepackage{array}
\usepackage[english]{babel}
\usepackage{makecell}
\usepackage{dsfont}
\usepackage{enumitem}
\usepackage{comment}
\usepackage{nameref}
\usepackage{color}
\usepackage[outdir=./]{epstopdf}
\usepackage[colorlinks = true, citecolor = blue, urlcolor =cyan]{hyperref}
\usepackage[figure]{hypcap}
\usepackage{multirow}
\usepackage{makecell}
\usepackage{xcolor}
\usepackage{mathtools}
\usepackage{mathdots}
\usepackage{tikz}      
\usetikzlibrary{knots}  
\usepackage[notransparent]{svg}
\usepackage[normalem]{ulem}
\usepackage{verbatim}

\newcolumntype{C}{>{$}c<{$}}
\AtBeginDocument{
\heavyrulewidth=.08em
\lightrulewidth=.05em
\cmidrulewidth=.03em
\belowrulesep=.65ex
\belowbottomsep=0pt
\aboverulesep=.4ex
\abovetopsep=0pt
\cmidrulesep=\doublerulesep
\cmidrulekern=.5em
\defaultaddspace=.5em
\tabcolsep=7pt
}
\usepackage{booktabs}
\usepackage{soul}

\definecolor{emerald}{rgb}{0.07, 0.53, 0.03}

\usepackage{braket}

\usepackage{amsbsy}

\usepackage{nicefrac} 

\usepackage{soul}   
\setstcolor{red} 
\setul{}{1.5pt}  

\usepackage{titlesec}
\titleformat{\paragraph}[runin]
  {\normalfont\normalsize\itshape}
  {}
  {0pt}
  {}
  [:]
\titlespacing*{\paragraph}
  {0pt}
  {1.5ex plus .5ex minus .5ex}
  {0.5em}

\makeatletter
\let\latex@addcontentsline\addcontentsline
\renewcommand{\addcontentsline}[3]{}
\makeatother

\tikzset{
	smallcrosspic/.style={
		baseline=0.0ex,
		x=0.6em,
		y=0.6em,
		strand/.style={line width=0.5pt,line cap=round},
		overcross/.style={
			double,
			line width=0.5pt,
			white,
			double=black,
			double distance=0.6pt
		}
	}
}

\tikzset{
	bigcrosspic/.style={
		baseline=0.0ex,
		x=1.2em,
		y=1.2em,
		strand/.style={line width=1.0pt,line cap=round},
		overcross/.style={
			double,
			line width=1.0pt,
			white,
			double=black,
			double distance=1.2pt
		}
	}
}

\newcommand{\cloop}{%
	\begin{tikzpicture}[smallcrosspic]
		\draw[strand] (0,1.0) .. controls (0.5,0.0) and (1.0,0.0) .. (1.5,1.0);
		\draw[strand] (0,0) .. controls (0.5,1.0) and (1.0,1.0) .. (1.5,0.0);
	\end{tikzpicture}%
}

\newcommand{\clines}{%
	\begin{tikzpicture}[smallcrosspic]
		\draw[strand] (0,1.0) .. controls (0.5,0.5) and (1.0,0.5) .. (1.5,1.0);
		\draw[strand] (0,0) .. controls (0.5,0.5) and (1.0,0.5) .. (1.5,0.0);
	\end{tikzpicture}%
}

\begin{document}

\title{Realization of a Quantum Topological Photon Pump}

\author{Qianao Yue}
\email{qyue@umd.edu}
\affiliation{Department of Physics, University of Maryland, College Park, MD 20742, USA}
\affiliation{Joint Quantum Institute, NIST/University of Maryland, College Park, Maryland 20742 USA}

\author{Bernardo Barrera}
\affiliation{Department of Physics, Boston University, 590 Commonwealth Avenue, Boston, Massachusetts 02215, USA}

\author{Martin Ritter}
\affiliation{Department of Physics, University of Maryland, College Park, MD 20742, USA}
\affiliation{Joint Quantum Institute, NIST/University of Maryland, College Park, Maryland 20742 USA}

\author{David M. Long}
\affiliation{Department of Physics, Stanford University, Stanford, California 94305, USA}

\author{David A. Lane}
\affiliation{Department of Physics, University of Maryland, College Park, MD 20742, USA}
\affiliation{Joint Quantum Institute, NIST/University of Maryland, College Park, Maryland 20742 USA}

\author{Anushya Chandran}
\affiliation{Department of Physics, Boston University, 590 Commonwealth Avenue, Boston, Massachusetts 02215, USA}

\author{Alicia J.  Koll\'ar}
\affiliation{Department of Physics, University of Maryland, College Park, MD 20742, USA}
\affiliation{Joint Quantum Institute, NIST/University of Maryland, College Park, Maryland 20742 USA}
\affiliation{Maryland Quantum Materials Center, Department of Physics, University of Maryland, College Park, MD 20742, USA}

\preprint{APS/123-QED}

\date{\today}

\begin{abstract}
Topological pumps transfer charge or energy at quantized rates determined solely by band topology, independent of the details of the control fields. 
Photon pumps operating on this principle could enable the robust preparation of non-classical states in a quantum cavity, even in the presence of control imperfections, but have not been directly observed. 
We present the first experimental observation of a topological photon pump, using a transmon qubit coupled to a microwave cavity. The pump operates in the quantum regime, pumping up from the vacuum state up to $\approx 7$ photons, and produces demonstrably non-classical cavity states for the first few cycles.
\end{abstract}
\maketitle











Non-trivial band topology produces discrete band invariants, which in turn lead to robust phenomena, such as quantized conductivity and protected edge states. In periodically driven systems, these invariants protect responses locked to the period of the cycle. In particular, winding numbers of the stationary (quasi-energy) states can protect quantized pumping, in which quantities such as charge or energy change by a discrete amount per period~\cite{Rudner:2013mz, Nathan:2015aa, Po:2016aa, Roy:2017aa}.
A canonical example is the one-dimensional Thouless pump~\cite{Thouless1983pump}, in which the periodic modulation of an insulating chain transports an integer number of particles from one end of the chain to the other in every cycle. Such spatially extended pumps have been experimentally realized in ultracold atoms, photonic waveguide arrays, and synthetic Rydberg dimensions~\cite{Cheng_topo_pump_metamaterial_2020,quasi_crystal_photonic_Thouless, fermion_Thouless_pump, Lohse2018, Thouless_pump_Lohse_2016, Thouless_pump_disordered_photonics_Cerjan,Jurgensen_nonlinear_Thouless_pump_2021,Mostaan_nonlinear_Thouless_soliton_atoms_2022,grinberg_robust_2020, walter_quantization_2023, nakajima_competition_2021, jurgensen_quantized_2023,SC_Thouless_2025, Wintersperger2020}. 

Topological photon pumps -- in which photons are pumped into (or out of) a cavity at a quantized rate -- have long been sought. These pumps can be used to robustly prepare non-classical photon states, such as Fock states with high photon numbers or Schr\"odinger cat states, which in turn are sensitive sensors~\cite{Giovannetti2011metrology,deng2024quantum}, or can be used to encode quantum information~\cite{joshi2021quantum}. The reverse pump operation can also perform on-demand cavity reset to the vacuum state.
However, despite several theoretical proposals~\cite{Martin:2017aa,Crowley:2019_classification,Nathan:2019_drivendiss,Long2022b} and a few proof-of-principle experiments~\cite{cQED_two_classical_drives_Malz,Boyers2020}, photon pumping has not been directly observed.

\begin{figure}
    \centering
    \includegraphics[width=\linewidth]{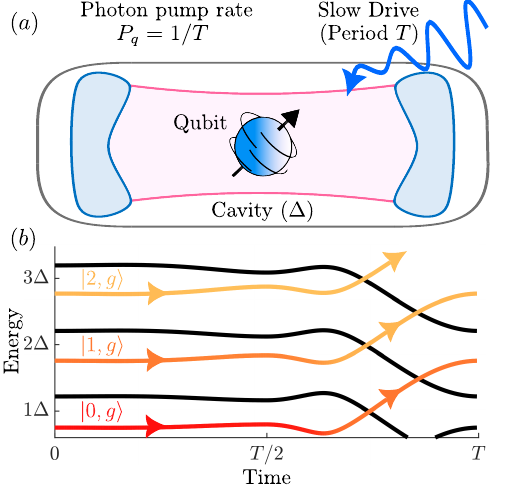}
    \caption{\emph{The Topological Photon Pump}. (a) A coupled qubit-cavity system is periodically driven by a slow external drive with period $T$. One photon enters the cavity per cycle, so that the average pump rate $P_q=1/T$. (b) The instantaneous spectrum vs time, showing spectral flow. Along each colored line, an $n$-photon state $\ket{n,g}$ is adiabatically transported one rung up the eigenstate ladder to an $(n+1)$-photon state $\ket{n+1,g}$ in one period; over multiple periods, the number of photons increases from $n=0$ to $n=n_\mathrm{max}$. The black lines correspond to depumping states.} 
    \label{fig:pumping_schematic}
\end{figure}

By driving a superconducting transmon qubit coupled to a microwave cavity~\cite{Long2022b,toolkitmartin} (Fig.~\ref{fig:pumping_schematic}(a)), we realize a topological photon pump and observe quantized photon pumping in the quantum regime, in which the pump initiates from the vacuum state of the cavity. 

The topological photon pump relies on the spectral flow, or winding, of the instantaneous spectrum of the joint cavity-qubit system (Fig.~\ref{fig:pumping_schematic}(b)). During one cycle of the external drive, the state $\ket{n,g}$, with $n$ intracavity photons and the qubit in its ground state, connects to the state $|n+1, g\rangle$, as shown by the colored lines. If the system follows the colored lines adiabatically, precisely one photon is pumped into the cavity each cycle, so that the average pump rate $P_q = 1/T$ is quantized.
Along the black lines, on the other hand, the state $\ket{n,e}$ transforms to $\ket{n-1,e}$, resulting in quantized depumping.

We demonstrate pumping starting from the vacuum state and achieving a maximum photon number of $\langle n\rangle \approx 7$. 
The photon statistics of the resulting cavity state are measurably sub-Poissonian up to $\langle n \rangle \approx 3$.
We characterize the phase diagram of the pump both analytically and experimentally, and demonstrate the robustness and quantization of the pump in the presence of imperfections.
The photon pump operates best in the large cavity-detuning regime, and is physically quite different from the small detuning pump discussed in previous proposals~\cite{Nathan:2019aa,Long2022b}.

\begin{figure*}[t]
    \centering
    \includegraphics[width=\textwidth]{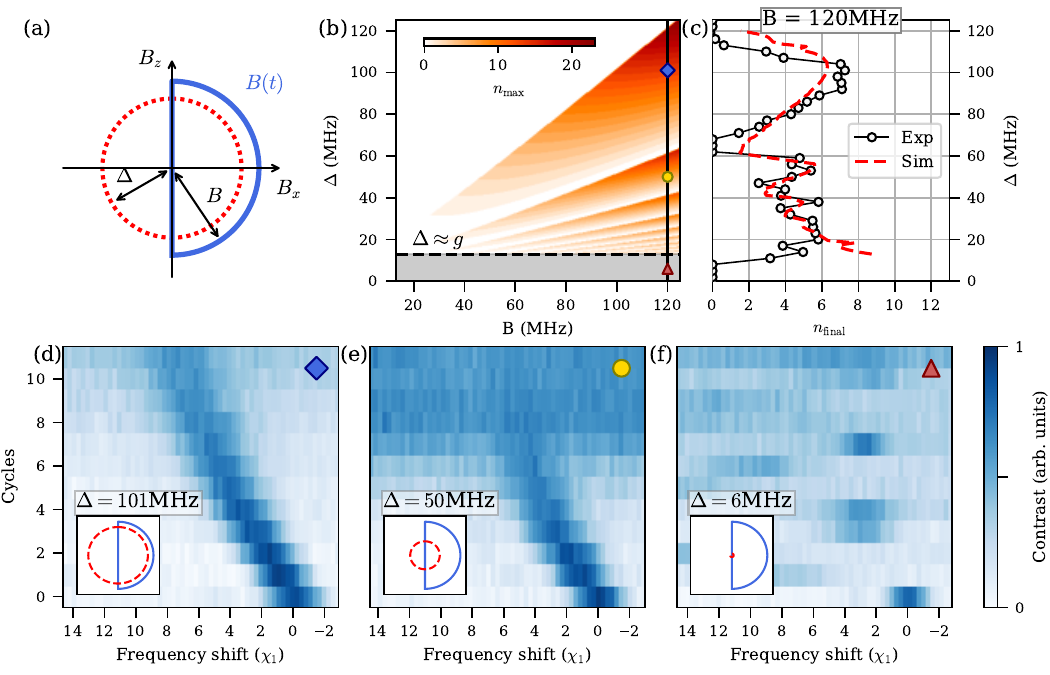}
    \caption{\textit{Phase Diagram of the Topological Pump.} 
    (a) Schematic of the effective time-dependent field $\vec{B}(t)$ (blue) relative to the ring $|\vec{B}(t)| = \Delta$.
    (b) The analytically computed maximum photon number $n_{\max}$ (color) as a function of $B$ and $\Delta$ for $g = 13$~MHz, neglecting cavity loss. 
    The resonances at $\Delta_\ell$, where $n_{\max} =0$ separate large regions of robust pumping, with the largest values of $n_{\max}\approx 20$ found in the range $\Delta_2< \Delta <\Delta_1$ (blue diamond).
    Panels (c-f) show data along the black line $B = 120$~MHz. 
    (c) Experimentally achieved largest photon number $n_{\rm{final}}$ over $11$ cycles (circles) and numerically simulated values accounting for cavity decay (red dashed line) versus detuning at $B = 120$~MHz (black line in (b)), showing good agreement. 
    (d,e,f) Spectroscopy of the AC Stark shift $n\chi_1$ on the qubit versus number of drive cycles for $B$ and $\Delta$ corresponding to the markers in (b).
    Systematic increase of the AC Stark shift (d,e) indicates topological pumping and corresponding increase in $n$.
    Drive protocol and detuning for each panel shown in inset.
    (f) Back-action dominated regime with no systematic pumping.
    Additional data are shown in Fig.~\ref{fig:detuning_sweep_sup} of the Supplemental Materials.
    }
    \label{fig:main_2}
\end{figure*}

\paragraph{The model} 
The system consists of a qubit which is coupled to a cavity with strength $g$ and is subject to a time-dependent effective magnetic field $\vec{B}(t)$ $(\hbar=1)$:
\begin{equation}
	H_{\mathrm{pump}}(t) = \Delta\, a^\dagger a + g(a^\dagger \sigma^- + a \sigma^+) + \frac{1}{2}\vec{B}(t)\cdot\vec{\sigma},
	\label{eq:JCH_rot}
\end{equation}
where $a^\dagger$ ($a$) is the photon creation (annihilation) operator, $\vec\sigma$ is the vector of Pauli operators acting on the qubit, and $\Delta = \omega_c - \omega_q$, the effective cavity frequency, is the detuning between cavity and the laboratory-frame qubit frequencies.
If $|\vec{B}(t)| = \Delta$, the qubit and cavity become resonant and a photon can be exchanged between them. 
Systematic pumping is achieved when this exchange does not average out over the cycle, even though $\vec{B}(t)$ is periodic.
Here we consider an effective field, shown in Fig.~\ref{fig:main_2}(a), which traces a semicircular trajectory of magnitude $B$ in the $(B_x,B_z)$ plane with rotation rate $\omega$ and period $T = 2\pi/\omega$:
\begin{equation}
\vec{B}(t) = 
\begin{cases}
      B\,  \left[ \sin(\omega t),0,-\cos(\omega t) \right] &  0 \leq t \leq T/2,\\
     B\,  \left[ 0,0,-\cos(\omega t) \right]  &   T/2 \leq  t \leq T.
\end{cases}
\end{equation}

The evolution of the instantaneous spectrum of $H_{\rm pump}(t)$ and the action of the pump in the regime of large $\Delta$, plotted in Fig.~\ref{fig:pumping_schematic}(b), are simply understood.
At $t=0$, the spectrum is composed of two ladders of photon states, one for each qubit state $\ket{g}, \ket{e}$. The energies are nearly constant during the semicircular part of the drive protocol ($0<t<T/2$), although the eigenstates continuously evolve from $\approx |n,g\rangle$ to $\approx |n, e\rangle$ (and vice-versa). During the vertical part of the drive protocol ($T/2\leq t\leq T$), the spectrum is that of the Jaynes-Cummings (JC) model. In particular, there is an avoided level crossing between states with the same polariton number ($\ket{n,e}$ and $\ket{n+1,g}$) at \(B_z(t) = \Delta\) and two exact level crossings between states of different polariton numbers  near  \(B_z(t) = 0\) and \(B_z(t) = -\Delta\) (seen as exact level crossings between colored and black lines in Fig.~\ref{fig:pumping_schematic}(b)). 
Thus, during the semicircular part of the drive protocol, only the qubit state flips; whereas, in the vertical part, this qubit excitation is transferred to the cavity, returning the qubit to $\ket{g}$.
For simplicity, Fig.~\ref{fig:pumping_schematic} shows the case where $\Delta \lesssim B \lesssim 2\Delta$; otherwise, there are additional exact level crossings.

In order that the system stays close to the eigenstates $H_{\rm{pump}}(t)$ at each $t$, the following hierarchy of energy scales is required: 
\begin{equation}
	  \kappa \ll \omega < g \ll \Delta,  B,
	\label{eq:hierarchy_scales}
\end{equation} 
where $\kappa$ is the cavity decay rate. The joint cavity-qubit system should evolve adiabatically through the avoided level crossing; this gives $\omega<g$. The criterion $\kappa \ll \omega$ ensures that several pump cycles occur before significant photon loss from the cavity. Lastly, $g \ll \Delta$ suppresses large variations of the cavity photon number because of the qubit state variation. These can destabilize the pump, as we discuss later.

\paragraph{Observation of photon pumping}
We realize Eqs.~\eqref{eq:JCH_rot},~\eqref{eq:hierarchy_scales} in a circuit QED platform consisting of a tunable-frequency transmon qubit coupled to a coplanar waveguide cavity~\cite{Blais_review,quantum_engineer_guide}.
Following the proposal in Ref.~\cite{Long2022b,toolkitmartin}, the effective field $\vec{B}(t)$ is synthesized through amplitude modulation of the Rabi drive together with flux modulation of the qubit frequency, and the Hamiltonian $H_{\rm pump}(t)$ [Eq.~(\ref{eq:JCH_rot})] is realized in a frame rotating at the drive carrier frequency, neglecting counter-rotating terms.
Our device, previously described in Ref.~\cite{toolkitmartin}, features a cavity with frequency $\omega_c = 2\pi \times 5.04$~GHz, cavity-qubit coupling $g = 2\pi \times 13$~MHz, a cavity decay time of $2 \ \mu\rm{s} = 1/ \kappa = 1/(2 \pi \times 84 \mbox{ kHz})$, and allows effective magnetic fields of up to $B = 2 \pi \times 125$~MHz. 
Factors of $2 \pi$ in angular frequencies are suppressed in the remainder of the manuscript.

The key experimental demonstration of topological photon pumping is shown in Fig.~\ref{fig:main_2}(d). Starting from the state $\ket{0,g}$ at $t= 0$, we observe a maximum photon number 
of approximately $7$
using $B = 120~\mathrm{MHz}$, detuning $\Delta = 7.7\, g = 101~\mathrm{MHz}$, and rotation rate $\omega = 5~\mathrm{MHz}$.
Cavity population $n(mT)$ is deduced from qubit spectroscopy, which measures the cavity-photon-induced AC Stark shift after the $m$th drive cycle $\chi(mT) = n(mT) \chi_1$, where $\chi_1$ is the 1-photon shift~\cite{Blais_review, quantum_engineer_guide}.
After completion of $m$ cycles, the 
control fields are turned off adiabatically, and the qubit is brought to a reference detuning of $140$~MHz below the cavity,
where a variable-frequency (spectroscopy) $\pi$-pulse attempts to invert the qubit population. 
(See Sec.~\ref{sec:experimental_realization} of Supplemental Materials for details.)
Maximum contrast occurs when the spectroscopy-pulse frequency matches the AC Stark shift, corresponding to dark regions in Fig.~\ref{fig:main_2}(d). 

Measured $n(mT)$ for $\omega$ between $1~\mathrm{MHz}$ and $10~\mathrm{MHz}$ are shown in Fig.~\ref{fig:omega_sweep_sup} of the Supplemental Materials.

\paragraph{Phase diagram of pumping}
Exploring the parameter space of the topological photon pump reveals three notable behaviors: robust pumping over a large range of parameters;  a nontrivial dependence of the maximum number of pumped photons $n_{\rm max}$ on $\Delta$, with sub-structure related to resonances;
and pronounced photon number variations that disrupt pumping in the small $\Delta$ regime, where topologically-protected photon pumping was previously predicted~\cite{Nathan:2019aa,Long2022b}. 

Pumping is robust over a large range of parameters because the spectral flow of the instantaneous eigenstates of $H_{\rm{pump}}$ is not fine-tuned.
For given $\Delta$ and external field amplitude $B$, non-trivial spectral flow starting from the state $\ket{0,g}$ requires: 
\begin{eqnarray}
    B > B_{\rm{min}}(\Delta) =&  \frac{\Delta^2 - g^2}{\Delta}, \label{eq:deltamax}\quad \mbox{ or,}\\
\Delta < \Delta_{\rm{max}}(B) =& \frac{B+\sqrt{B^2+4g^2}}{2}.  \label{eq:bmin}
\end{eqnarray}
Spectral flow persists until the state $\ket{n_{\rm{max}}, g}$ is reached, when pumping and depumping states become nearly degenerate. The maximum theoretically achievable photon number is thus $n_{\rm{max}}$.  (See Eq.~\eqref{eq:maximum_photon_number} of the Supplementary Materials for an analytical formula of $n_{\rm max}$ in the $\omega \rightarrow 0$ limit.)
Fig.~\ref{fig:main_2}(b) shows the analytically-computed phase diagram as a function of $B$ and $\Delta$, and values of $n_{\rm max}$ as large as $20$ are predicted at large $B, \Delta$.

The triangular region that satisfies Eqs.~\eqref{eq:bmin}, \eqref{eq:deltamax} and $\Delta \gtrsim g$ is broadly the pumping regime. In more detail, however, it is marked by a series of sharp resonances where $n_{\rm max}$ drops to zero because the state $\ket{0,g}$ is degenerate with the $(\ell+1)$-polariton state $\approx\ket{\ell,e}$ at the start of the protocol, for $\ell = 1,2, \ldots$.
These resonances occur at specific values of the detuning:
\begin{equation}
	\Delta_\ell = \frac{B+\sqrt{B^2 + 4g^2\ell}}{2\ell}, \qquad \ell = 1,2,3,\dots
    \label{eq:critical_detunings}
\end{equation}
where $\Delta_{\rm{max}} = \Delta_1$.
Thus, regions enclosed by the curves $\Delta =\Delta_\ell$ and $\Delta = \Delta_{\ell+1}$ are the range of parameters for which photon pumping up to highly excited states is possible. A full discussion of the phase diagram and the possible topological classes can be found in Sec.~\ref{sec:topological_classification} of the Supplemental Materials.

Experimentally, the largest photon number is limited either by the theoretical upper bound $n_{\rm max}$ or by cavity loss.
For example, the pump configuration shown in Fig.~\ref{fig:main_2}(e) with $\Delta = 3.8\, g = 50~\mathrm{MHz}$ is $n_{\max}$-limited. 
For the first few cycles, photons are transferred into the cavity at the rate $P_q=1/T$, and the cavity-qubit system evolution is adiabatic. 
Therefore, the qubit maintains polarization after each pump cycle, and the spectroscopy contrast remains large. As $n(mT)$ reaches $n_{\max}$, the system undergoes a diabatic transition and the qubit loses polarization, marked by a sharp loss in spectroscopy contrast. 
In contrast, the pump configuration shown in Fig.~\ref{fig:main_2}(d) with $\Delta = 7.7\, g = 101~\mathrm{MHz}$ is $\kappa$-limited. Here, the qubit never loses polarization, and the spectroscopy contrast remains large; however, the frequency shift of the qubit, and hence the average photon number in the cavity, shows signs of saturation after the eighth cycle due to cavity loss. In this case, the cavity photon number saturates at a value well below $n_{\max}=14$. 
Extended data, including at $\Delta \approx \Delta_\ell$, is shown in Fig.~\ref{fig:detuning_sweep_sup} of the Supplemental Materials.
 
Fig.~\ref{fig:main_2}(c) shows the detuning dependence of the largest photon number $n_{\rm{final}}$ achieved by the pump over $11$ cycles and the numerical simulation of $n_\mathrm{final}$ in the presence of cavity loss for $B = 120$~MHz.
The agreement is good; in particular, the sharp dips in the maximum photon number occur at the same values of $\Delta$. 
However, the two curves deviate when $\Delta \lesssim g$, which we now discuss. 

When $\Delta \lesssim g$, the qubit strongly affects the photon number of the cavity during the semicircular part of the protocol. This back-action can disrupt pumping. Consider for concreteness $t=T/4$, when $\vec{B}(t) = B \hat{x}$. For $g \ll B$, the qubit is approximately polarized along $x$, which creates an effective classical field on the cavity that displaces the state of the cavity to photon number $\approx g/\Delta \gtrsim 1$. The average photon number thus significantly varies with the qubit state in a cycle, with the scale of the variation exceeding the systematic change of $P_q T=1$ predicted for the topological pump. Furthermore, for moderate values of $B$, the average photon number can reach $n_\mathrm{max}$ within a cycle, in which case topological pumping from the vacuum state does not occur at all. (See Sec.~\ref{sec: pump breakdown} of the Supplemental Materials for details.)
The data in Fig.~\ref{fig:main_2}(f) taken at $\Delta = 0.5g = 6~\mathrm{MHz}$ clearly shows the predicted large variations of $n(mT)$ at small detunings. 

In summary, although the system can be a topological pump at low $\Delta$ and sufficiently high $B$~\cite{Yuan2018ringresonator,Nathan2022weyl,Lantagne-Hurtubise2024graphene,Luneau2022power,Long2021:class,Nathan2020b,Schwennicke2022enantioselective}, the regime is neither ideal for observing quantized photon pumping from the vacuum, nor for preparing non-classical cavity states (see below).

\begin{figure}[!t]
    \centering
    \includegraphics[width=\linewidth]{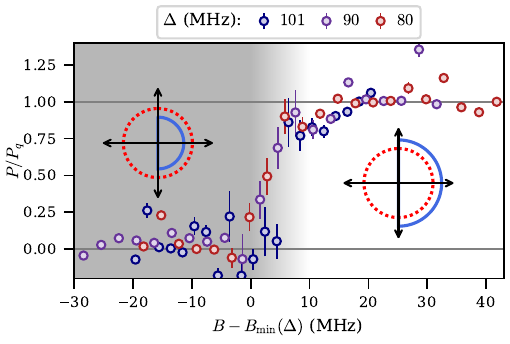}
    \caption{
    \textit{Quantization of Pump Rate.} Photon pump rate $P/P_q$ extracted from the first five cycles versus $B$ for $\Delta = 101$, $90$, $80$~MHz and $\omega=5$~MHz. Error bars indicate the uncertainty of the fit to $n(mT) = (P/\kappa) \times (1 - e^{-\kappa mT})$. We observe $P \approx 0$ for $B<B_{\rm min}(\Delta)$ (gray shaded region) in the trivial regime, and $P/P_q \approx 1$ deep in the topological regime $B>B_{\rm min}(\Delta) + 10\,$MHz. Insets show the drive protocols in the two regimes. Sample spectroscopy data corresponding to $\Delta=80$~MHz are shown in Fig.~\ref{fig:b_sweep_sup} of the Supplemental Materials.
    }
    \label{fig:main_3}
\end{figure}

\paragraph{Robust quantization of the pump rate}
Topological effects are characterized by insensitivity to small perturbations.
We show that the experimentally-observed photon pump rate $P$ matches the ideal quantized pump rate $P_q$ in the topological regime, independent of $B$ and $\Delta$.

In the presence of a finite cavity loss rate $ \kappa$, a constant pump rate $P$ results in $n(mT) \approx( P/\kappa) \times (1 - e^{-\kappa mT})$ after $m$ cycles. 
We determine $P$ from the observed $n(mT)$ for the first $5$ cycles and the known value of $\kappa$. 
Extracted values of $P$ versus $B$ for $\Delta = 80$, $90$, and $101$ MHz are shown in Fig.~\ref{fig:main_3}.
In the trivial regime where $B < B_{\rm{min}}(\Delta)$, $P\approx 0$. At $B = B_{\rm{min}}(\Delta)$, instantaneous gaps vanish, and the system evolution cannot be adiabatic for nonzero $\omega$, resulting in non-quantized pumping when $B - B_{\rm{min}} \lesssim \omega = 5$~MHz. We observe $P \approx P_q$ everywhere else in the topological regime, independent of the value of $\Delta$ or $B$. 
See Sec.~\ref{sec:quantized pumping} of the Supplemental Materials for extended data and characterization of $n(mT)$.

The photon pump is robust under a restricted class of deformations of the drive protocol. As shown in Fig.~\ref{fig:pumping_schematic}(b), spectral winding relies on exact level crossings between pumping and depumping states. Along the vertical stroke of the protocol, the transverse field vanishes, which protects exact crossings by polariton-number conservation in the Jaynes-Cummings model. Consequently, the spectral flow is stable under arbitrary deformations of the semicircular segment, provided the vertical stroke is left unchanged. By contrast, transverse fields applied during the vertical stroke generically turn the exact crossings into avoided level crossings, so adiabatic following no longer produces spectral flow. Due to our implementation of $H_{\rm{pump}}$ in a rotating frame, in which the \emph{effective} transverse magnetic field is induced by a microwave drive, ambient transverse-field noise is highly suppressed. Direct laboratory-frame implementations of $H_{\rm{pump}}$ would not, in general, benefit from such intrinsic noise protection.

The quantization of the pump rate can be alternatively understood from the Chern number of the qubit states as a function of the cavity and external drive phases; see Sec.~\ref{sec:chern_numbers} of the Supplementary Materials.
Experimental evidence for other characteristic signatures of topology, including reversal of the photon current when the drive direction is reversed, is presented in Sec.~\ref{sec:chirality} of the Supplementary Materials.

\begin{figure}[t]
    \centering
    \includegraphics[width=\linewidth]{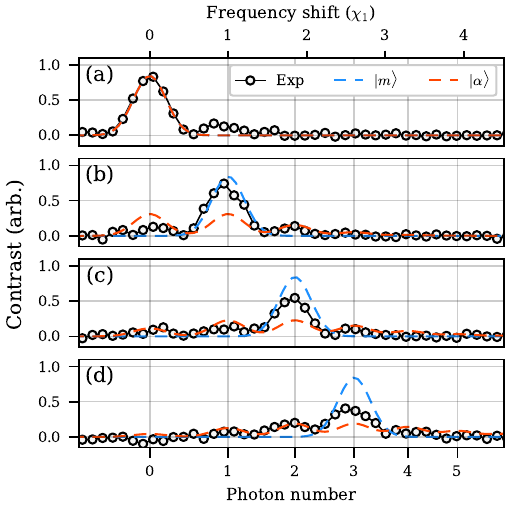}
    \caption{
    \textit{Fock-State Pumping.}
    (a-d) Photon-number-resolved qubit spectroscopy after $m=$ 0-3 pump cycles, showing experimental data (white circles), coherent-state distributions with $|\alpha|^2=m$ (red), and ideal $m$-photon Fock states (blue).
    The upper (lower) axis indicates integer multiples of the single-photon shift $\chi_1$ (the exact qubit frequency shifts corresponding to $m$ photons).
    Despite cavity decay, the measured photon-number distributions retain strong overlap with the 1, 2, and 3-photon Fock states.
    }
    \label{fig:main_4}
\end{figure}

\paragraph{Production of quantum states of light}
The photon pump generates non-classical cavity states from vacuum, as each ideal pump cycle takes any Fock state $\ket{n}$, including the vacuum, to $\ket{n+1}$. 
Unlike a conventional displacement operation, which produces coherent states with Poissonian photon statistics when acting on the vacuum, we experimentally observe strongly sub-Poissonian cavity states generated by the pump.
This behavior contrasts with previous proposals for the small-$\Delta$ pump~\cite{Long2022b}, which can amplify a starting Fock state with a few photons only at special rephasing points in time.

To characterize the states produced by the pump, qubit spectroscopy is carried out at a closer reference detuning of 40 MHz. This enhances the photon-number-dependent qubit splitting, so that individual Fock states become spectroscopically resolvable~\cite{Blais_review,quantum_engineer_guide}. Starting from vacuum, we observe a cavity state with strong overlap with the single-photon Fock state after the first pump cycle, as shown in Fig.~\ref{fig:main_4}(a,b). Subsequent pump cycles generate higher Fock states, while cavity loss transfers some population back toward lower photon-number states. We observe appreciable sub-Poissonian character in the pumped cavity states up to $\langle n \rangle =3$, as shown in Fig.~\ref{fig:main_4}(c,d). In the present device, cavity decay limits the fidelity of the generated Fock states, while nonlinearities in the photon-number-resolved qubit spectrum constrain the highest detectable Fock state. As superconducting microwave cavities with photon lifetimes exceeding a millisecond have already been demonstrated in both planar and 3D geometry~\cite{Spiralcavity2025,SchoelkopfHighQ2024,MilulHighQ3D}, the present limitations are not fundamental and substantially larger Fock states should be experimentally accessible.

\paragraph{Outlook}
We report the first experimental realization of a topological photon pump. The pump operates at a quantized rate for a wide range of parameters and  
produces Fock states when initialized from the vacuum.
Unlike fine-tuned methods that leverage optimal control~\cite{SchoelkopfOptimalControl} or photon-number-resolved gates~\cite{MartinisOptimalControl, SchoelkopfSNAP, Blais_review}, the quantized pumping and depumping are independent of the cavity state and are robust to imperfections and variations of the control fields.

The performance of the pump can be enhanced though straightforward experimental improvements
that reduce cavity loss and suppress non-adiabatic processes.
Cavity loss can be greatly reduced through improved fabrication or different cavity geometries ~\cite{Spiralcavity2025,SchoelkopfHighQ2024,MilulHighQ3D}, while non-adiabaticity can be suppressed by an additional lossy mode that selectively removes population from the depumping photon ladder~\cite{dissipation_paper_PRX}.
Thus, the topological pump presented here can readily be extended to support faster and higher-fidelity pumping, thereby producing larger and higher-quality Fock states.

Topological photon pumping could be a versatile and robust resource for quantum computing and sensing. By preparing the state of the qubit in a superposition of $|g\rangle$ (pumping) and $|e\rangle$ (depumping) states, desired superpositions of Fock states with very different photon numbers in the cavity can be achieved~\cite{Nathan:2019_drivendiss}, independent of the details of the control fields. Such cat states are resources for Heisenberg-limited quantum sensing and bosonic error correcting codes~\cite{Giovannetti2011metrology,deng2024quantum,joshi2021quantum}. Reversing the pump could also rapidly reset high-$Q$ bosonic modes independent of their initial state, a capability necessary for information processing with cavity modes. Adapting this protocol to optical cavities coupled to atomic or Rydberg ensembles could similarly generate on-demand large-$N$ photonic Fock states of optical photons---a long-standing goal in cavity QED.

\begin{acknowledgments}
We thank Maya Amouzegar for assistance with device fabrication, Benjamin Cochran for contributions to device modeling, and IBK Adisa for assistance with early measurements.

This work was supported by ARL (Grants no. W911NF-19-2-0181 and W911NF-17-S-0003), the University of Maryland, AFOSR (Grants FA9550-21-1-0129, FA9550-24-1-0121, and No. FA9550-20-1-0235), and the NSF (QLCI grant OMA-2120757). 
MR received support from the National Science Foundation (PFC at JQI Grant No. PHY-1430094), the LPS graduate fellowship, and ARCS. 
QY received support from AFOSR Grant No. FA9550-21-1-0129 and AFOSR Grant No. FA9550-24-1-0121.
DAL received support from AFOSR Grant No. FA9550-24-1-0121.
QY and DAL received additional support from NSF career (Grant No. PHY2047732) and ARO Grant No. W911NF-24-20240.
DML was supported by a Stanford Q-FARM Bloch postdoctoral fellowship, and the US Department of Energy, Office of Science (Award No.\ DE-SC0019380).
BB received support from AFOSR Grant No. FA9550-21-1-0342 and AFOSR Grant No. FA9550-24-1-0121.

\end{acknowledgments}

\bibliographystyle{apsrev4-2}
\bibliography{refs}

\clearpage
\onecolumngrid

\makeatletter
\let\addcontentsline\latex@addcontentsline
\makeatother

\setcounter{figure}{0} 
\setcounter{equation}{0}
\setcounter{section}{0}
\setcounter{tocdepth}{2}

\renewcommand{\figurename}{Figure} 
\renewcommand{\thefigure}{S\arabic{figure}} 
\renewcommand{\thetable}{S\arabic{table}} 
\renewcommand{\theequation}{S\arabic{equation}} 
\renewcommand{\thesection}{S\arabic{section}}

\renewcommand{\theHfigure}{S\arabic{figure}}
\renewcommand{\theHequation}{S\arabic{equation}}
\renewcommand{\theHsection}{S\arabic{section}}

\begin{center}
    \Large Supplemental Materials:\\
    Realization of a Quantum Topological Photon Pump
\end{center}

\tableofcontents

\vskip 0.6in

\clearpage
\section{Bidirectionality of the Photon Pump}
\label{sec:chirality}
\begin{figure*}[t]
    \centering
    \includegraphics{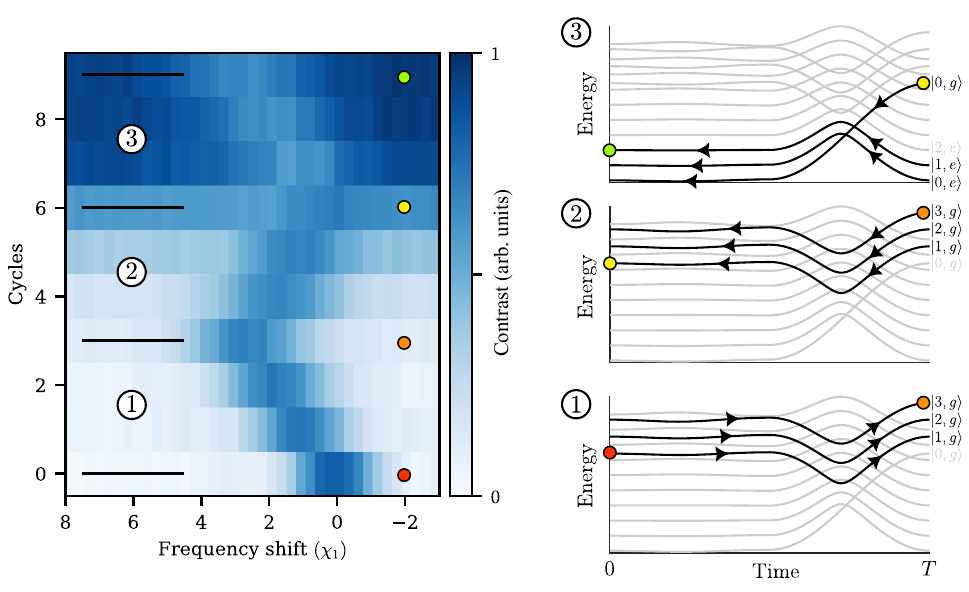}
    \caption{\textit{Bidirectionality of the Photon Pump.} Spectroscopy of the AC Stark shift on the qubit versus number of drive cycles showing bidirectionality of the photon pump with $\omega=5$ MHz, field amplitude $B=120$ MHz, and detuning $\Delta=20$ MHz. The experiment shows three regimes: 
    (1) photon pumping, in which the effective field traces a counterclockwise semicircular trajectory in the $(B_x,B_z)$ plane with $\vec{B}_{\mathrm{counter}}(t)$ in Eq.~\ref{counterclockwise};
    (2) photon depumping, with clockwise $\vec{B}_{\mathrm{clock}}(t)$ in Eq.~\ref{clockwise};
    and (3) photon pumping, due to the depumping state in regime (2) adiabatically evolving into the pumping state when the cavity is empty.
    Corresponding spectral flow of the instantaneous eigenstates of $H_{\mathrm{pump}}$ is shown on the right panels. Color markers indicate initial and final states in regimes (1), (2), and (3). Corresponding spectroscopy data are labeled with the same color markers on the left.}
    \label{fig:zig_zag}
\end{figure*}
A characteristic feature of the topological photon pump is that the sign of the photon current can be directly controlled by the direction of the drive protocol. Consider the semicircular effective field discussed in the main body of the manuscript. The field can trace out either a counterclockwise or clockwise trajectory in the $(B_x, B_z)$ plane with the following forms:
\begin{equation}
\vec{B}_{\mathrm{counter}}(t) = 
\begin{cases}
      B\,  \left[ \sin(\omega t),0,-\cos(\omega t) \right] &  0 \leq t \leq T/2,\\
     B\,  \left[ 0,0,-\cos(\omega t) \right]  &   T/2 \leq  t \leq T,
\end{cases}\label{counterclockwise}
\end{equation}
and 
\begin{equation}
\vec{B}_{\mathrm{clock}}(t) = 
\begin{cases}
     B\,  \left[ 0,0,-\cos(\omega t) \right]  &  0 \leq t \leq T/2,\\
     B\,  \left[ \sin(\omega t),0,-\cos(\omega t) \right]  &   T/2 \leq  t \leq T.
\end{cases}\label{clockwise}
\end{equation}
For the protocol using $\vec{B}_{\mathrm{counter}}(t)$, initializing the qubit in $\ket{g}$ or $\ket{e}$ puts the system into the pumping or depumping states that wind up or down the photon ladder, respectively. For example, consider the system initialized in the pumping state $\ket{n,g}$. During the semicircular part of the drive protocol ($0<t<T/2$), the system adiabatically rotates to $\ket{n,e}$.
Then, during the vertical part of the protocol ($T/2<t<T$) a photon is transferred to the cavity from the qubit. We define the following notation for this process:
\begin{equation}
	\ket{n,g} \xrightarrow{\text{circ}} \ket{n,e} \xrightarrow{\text{vert}} \ket{n+1,g},\label{Eq:CounterStatesinPump}
\end{equation}
where $\xrightarrow{\text{circ}}$ and $\xrightarrow{\text{vert}}$ represent the system adiabatically following along the semicircular and vertical part of the protocol. Similarly, initializing in the depumping state (for $n>0$), the system  winds down the photon ladder through the adiabatic evolution:
\begin{equation}
	\ket{n,e} \xrightarrow{\text{circ}} \ket{n,g} \xrightarrow{\text{vert}} \ket{n-1,e}.\label{Eq:CounterStatesoutPump}
\end{equation}

Conversely, for the protocol using $\vec{B}_{\mathrm{clockwise}}(t)$, initializing the qubit in $\ket{g}$ or $\ket{e}$ puts the system into the depumping or pumping states, respectively:
\begin{equation}
	\ket{n,g}\xrightarrow{\text{vert}} \ket{n-1,e}  \xrightarrow{\text{circ}} \ket{n-1,g},\label{Eq:StatesoutPump}
\end{equation} 
when the qubit is initialized in $\ket{g}$; and 
\begin{equation}
	\ket{n,e}\xrightarrow{\text{vert}} \ket{n+1,g}  \xrightarrow{\text{circ}} \ket{n+1,e},\label{Eq:CounterStatesinPump}
\end{equation} 
when the qubit is initialized in $\ket{e}$.

We directly observe the reversal of photon current upon changing the direction of the drive protocol in Fig.~\ref{fig:zig_zag}. Starting from $|0,g\rangle$, we run $\vec{B}_{\mathrm{counter}}(t)$ and the pump transfers approximately one photon into the cavity per cycle. After three cycles, the cavity occupation reaches $n \approx3$, as evidenced by the linear increase in measured AC Stark shift on the qubit. After the third cycle, the direction of the pump is reversed. Now, the state originally corresponding to photon pumping becomes a depumping state. The system then transfers photons out of the cavity, and after three cycles the system approximately returns to $|0,g\rangle$, reaching the lower topological boundary. Because photons cannot be removed from $\ket{0,g}$, continuing in this direction causes the system to wind back onto $\ket{0,e}$ (a pumping state of the clockwise direction) adiabatically
\begin{equation}
	\ket{0,g}\xrightarrow{\text{vert}} \ket{0,g}  \xrightarrow{\text{circ}} \ket{0,e}.\label{Eq:Bounceback}
\end{equation}
This transition is directly observed through a reversal of the spectroscopy contrast, indicating that the qubit now occupies $|e\rangle$ at the end of each cycle rather than $|g\rangle$. The zig-zag pumping -- depumping -- pumping, shown in Fig.~\ref{fig:zig_zag}, directly demonstrates the controllable directionality of the photon pump. The drive protocol for Fig.~\ref{fig:zig_zag} uses a field amplitude of $B=120$ MHz, a detuning of $\Delta=20$ MHz, and a rotation rate of $\omega=5$ MHz.

\section{Theoretical Model for the Large-$\Delta$ Pump}
\label{sec:theory_appendix}
This section develops a theoretical description of the topological photon pump in the large-$\Delta$ regime. It is organized as follows:
Sec.~\ref{sec:semi_classical_model} introduces a simple model for the pump consisting of a qubit driven by two tones, which is obtained from Eq.~\eqref{eq:JCH_rot} by treating the cavity semi-classically. Sec.~\ref{sec:topological_classification} classifies the different regimes of operation of the semi-classical pump in terms of the spectral flow of its Floquet quasi-energy spectrum. We give analytical expressions for the boundaries separating these regimes and connect them to the phase diagram of the quantum pump shown in Fig.~\ref{fig:main_2}(b) of the main text. Finally, Sec.~\ref{sec:chern_numbers} provides an alternative discussion of the topological classification in terms of the Chern number of the adiabatic state followed by the qubit.

\subsection{Semi-Classical Model}
\label{sec:semi_classical_model}
	
\paragraph{Setup} Much of the physics of the large-$\Delta$ pump can be understood within a semi-classical description of the cavity. Taking the expectation value of $H_{\rm pump}$ in a coherent state of the cavity $\ket{\alpha} = \ket{\sqrt{n}e^{-i\theta_2}}$, so that $g(\langle a^\dagger\rangle\sigma^- + \langle a\rangle\sigma^+) = g\sqrt{n}(\cos\theta_2\sigma_x+\sin\theta_2\sigma_y)$, yields the following effective Hamiltonian for the qubit alone,
\begin{equation}
	H_{\rm pump}^{\rm sc}\big(\theta_1(t),\theta_2(t)\big) = \frac{1}{2}\vec{B}\big(\theta_{1}(t)\big)\cdot\vec{\sigma} + \frac{1}{2}\vec{B}_{\rm cav}\big(\theta_2(t)\big)\cdot \vec{\sigma},
    \label{eq:H_pump_supp}
\end{equation}
where,
\begin{equation}
	\vec{B}(\theta_1) = B\left(\max [0,\sin\theta_1]\,\hat{x} - \cos\theta_1\,\hat{z}\right), \qquad 
	\vec{B}_{\rm cav}(\theta_2) = 2g\sqrt{n}\left(\cos\theta_2\,\hat{x}+\sin\theta_2\,\hat{y}\right),
    \label{eq:B_fields_supp}
\end{equation}
are the external and semi-classical cavity fields, respectively.
The norm of the semi-classical cavity field is $|\vec{B}_{\rm cav}| = 2g\sqrt{n}$. The external drive and cavity angles evolve in time as $\theta_1(t)=\omega t + \phi_1$ and $\theta_2(t)=\Delta t+\phi_2$. The period of the external drive is denoted as $T=2\pi/\omega$. For simplicity, we take the detuning to be positive: $\Delta>0$.

In the small-$\Delta$ regime, defined by the hierarchy of scales $\omega,\Delta\ll g\ll B$, the qubit adiabatically follows the total effective field $\vec{B}_{\rm eff}(\theta_1,\theta_2)=\vec{B}(\theta_1)+\vec{B}_{\rm cav}(\theta_2)$. The pumping rate is then set by the Chern number of the adiabatic state followed by the qubit, $\ket{\psi_\pm(\theta_1,\theta_2)} =\ket{\pm\hat{B}_{\rm eff}(\theta_1,\theta_2)}$, over the torus $(\theta_1,\theta_2)\in[0,2\pi)^2$~\cite{Martin:2017aa, Crowley:2019_classification}.     

In the large-$\Delta$ regime, defined by the hierarchy of scales $\omega\ll g\ll B,\Delta$, adiabaticity with respect to the instantaneous $\vec{B}_{\rm eff}$ breaks down. As we show below, the qubit instead follows a dressed adiabatic state $\ket{\widetilde{\psi}_\pm(\theta_1,\theta_2)}$. Like in the small-$\Delta$ case, the pumping rate is set by the Chern number of $\ket{\widetilde{\psi}_\pm(\theta_1,\theta_2)}$ (see discussion in Sec.~\ref{sec:chern_numbers}). 

\paragraph{Floquet States and Quasi-Energies}
The dressed adiabatic states $\ket{\widetilde{\psi}_\pm(\theta_1,\theta_2)}$ are found by solving the Floquet problem over the fast cavity angle $\theta_2$, at fixed slow drive angle $\theta_1$. The relevant Hamiltonian is,
\begin{equation}
	H_{\rm pump}^{\rm sc}\big(\theta_2(t);\theta_1\big) = \frac{1}{2}\vec{B}(\theta_1)\cdot\vec{\sigma} + \frac{1}{2}\vec{B}_{\rm cav}\big(\theta_2(t)\big)\cdot\vec{\sigma},
\end{equation}
where the semicolon denotes that $\theta_1$ is treated as a fixed parameter, and consequently $H_{\rm pump}^{\rm sc}(\theta_2(t);\theta_1)$ is periodic in time,
\begin{equation}
	H_{\rm pump}^{\rm sc}\big(\theta_2(t+T_2);\theta_1\big) = H_{\rm pump}^{\rm sc}\big(\theta_2(t);\theta_1\big), \qquad T_2 = \frac{2\pi}{\Delta}.
\end{equation}
Due to Floquet's theorem, there exist special time-periodic (up to a phase) solutions of the Schr\"odinger equation of the form,
\begin{align}
    \ket{\Psi_\pm(t;\theta_1)} = e^{-i\varepsilon_\pm(\theta_1)t}\ket{\widetilde{\psi}_\pm(\theta_1,\theta_2(t))},
	\qquad 
	i\partial_t  \ket{\Psi_\pm(t;\theta_1)} = H_{\rm pump}^{\rm sc}\big(\theta_2(t);\theta_1\big)\ket{\Psi_\pm(t;\theta_1)}.
    \label{eq:Floquet_solutions}
\end{align}
Eq.~\eqref{eq:Floquet_solutions} defines the dressed adiabatic states $\ket{\widetilde{\psi}(\theta_1,\theta_2)}$. The separation of frequency scales $\omega\ll\Delta$ allows a qubit that is initialized in $\ket{\widetilde{\psi}(\theta_1(0),\theta_2(0)}$ at $t=0$ to adiabatically follow the instantaneous state $\ket{\widetilde{\psi}(\theta_1(t),\theta_2(t)}$, so long as the Floquet quasi-energy gap $|\varepsilon_+(\theta_1) - \varepsilon_-(\theta_1)|$ is sufficiently large as compared to $\omega$. 

A similar analysis was employed in Ref.~\cite{Kolodrubetz2018} for the topological Floquet-Thouless energy pump. There, the authors constructed a topological invariant from the winding number of the micromotion (equivalent to $\theta_2$ here) of the Floquet states at some fixed value of an adiabatic parameter (equivalent to $\theta_1$ here).
	
The Floquet problem is exactly solvable along the vertical part of the protocol where $\vec{B}=B_z\hat{z}$. The exact quasi-energies along this segment are given by,
\begin{equation}
	\varepsilon_\pm(B_z)
	= \pm \frac{1}{2}\sqrt{(B_z-\Delta)^2 + B_{\rm cav}^2}-\frac{1}{2}\Delta
	\qquad (\mathrm{mod}\,\Delta). 
	\label{eq:exact_quasi_energies}
\end{equation}
Eq.~\eqref{eq:exact_quasi_energies} can be obtained by performing a rotating frame transformation $U(t) = \exp\left(-i\theta_2(t)\sigma_z/2\right)$, after which the Hamiltonian becomes time-independent in the rotating frame. 

\subsection{Phase Diagram}	
\label{sec:topological_classification}
\begin{figure}[t]
\begin{center}
	\includegraphics[width=\linewidth]{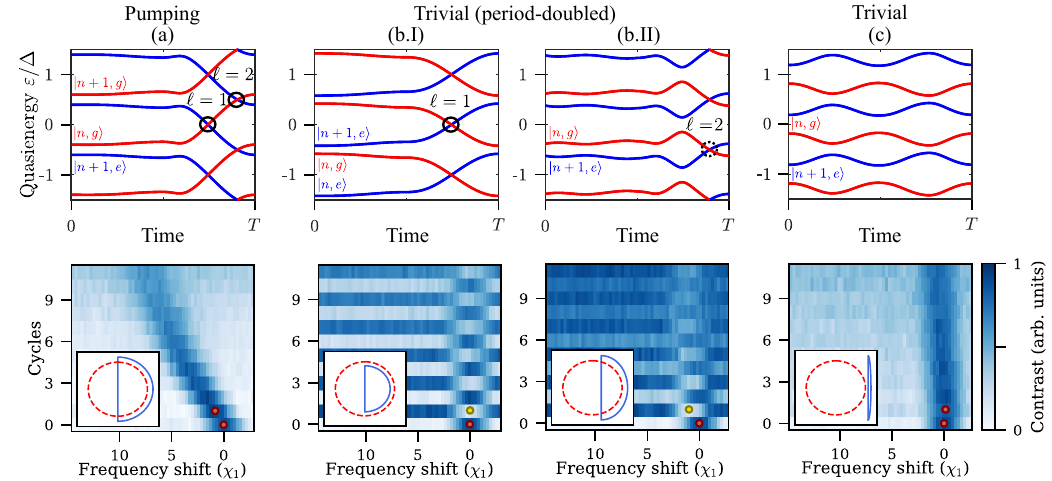}
\end{center}
\vspace{-0.6cm}
\caption{
\textit{The three regimes of the photon pump} are (a) topologically non-trivial with quantized photon pump rate $P_q$, (b) trivial with dynamics at twice the period of the drive $2T$, and (c) trivial with dynamics at the same period as the drive $T$. 
\textbf{Upper row}: Floquet quasi-energies of the related semiclassical model. (a) Pumping regime. Adiabatic evolution over one period maps a Floquet state onto itself, but displaced by one Floquet zone. When the cavity is quantized, this corresponds to the absorption of one photon per cycle: $|n,g\rangle \to |n+1,g\rangle$. 
(b) Period-doubled trivial regime. Adiabatic evolution over one period maps a Floquet state onto the Floquet state of opposite type. In example (b.I), the two Floquet states differ by $\ell=1$ Floquet zones, whereas in example (b.II) they differ by $\ell=2$ Floquet zones. When the cavity is quantized, (b.I) alternates between the two states $\approx\ket{n,g}$ and $\approx\ket{n,e}$ that differ by $\ell=1$ polariton numbers, whereas (b.II) alternates between the two states $\approx |n,g\rangle$ and $\approx |n+1,e\rangle$ which differ by $\ell=2$ polariton numbers. 
(c) Trivial regime. Adiabatic evolution over one period maps a Floquet state onto itself, in the same Floquet zone. When the cavity is quantized, there is no absorption of photons per cycle: $|n,g\rangle \to |n,g\rangle$.
In all panels, the magnitude of the semi-classical cavity field is set to be $B_{\rm cav}=2g\sqrt{n}$ with $n=1$.
\textbf{Lower row}: Experimental data showing spectroscopy of the AC Stark shift on the qubit versus number of drive cycles for each scenario. White background corresponds to state $\approx \ket{n,g}$ and blue to state $\approx \ket{n,e}$. Red and yellow markers denote the location of the qubit transition in the two cases. Drive protocol is shown in the inset. 
}
\label{fig:regimes} 
\end{figure}

Quantized energy transfer between the cavity and the external drive occurs when the Floquet quasi-energies $\varepsilon_\pm(\theta_1)$ exhibit spectral flow. Representative quasi-energy spectra in distinct regimes are shown in the upper panels of Fig.~\ref{fig:regimes}, with corresponding experimental data shown in the lower panels. The color of each quasi-energy branch indicates the character of the respective Floquet state at $t=0$: red branches are ground-state-like $\ket{g}$, whereas blue branches are excited-state-like $\ket{e}$. As a function of the dimensionless ratios $\Delta/B$ and $B_{\rm cav}/B$, the system exhibits three distinct topological regimes, labeled (a), (b), and (c).

\begin{enumerate}[label=(\alph*)]
	\item \textit{Pumping regime}. Adiabatic evolution over one external drive period $T$ maps a Floquet state onto itself, but displaced by one Floquet zone. The quasi-energies can be sorted into pumping and depumping branches satisfying,
	\begin{equation}
		\int_0^{2\pi}{\rm d}\theta_1 \,
		\frac{\partial \varepsilon_\pm(\theta_1)}{\partial \theta_1}
		= \pm \Delta.
	\end{equation}
	As a result, the cavity and external drive exchange a quantized amount of energy each period. 
    
    An example of the quasi-energy spectrum in the pumping regime is shown in the upper panel of Fig.~\ref{fig:regimes}(a). A pumping line (winding upward, colored red) crosses two depumping lines (winding downward, colored blue) per period. After one period, a pumping line adiabatically connects to another pumping line shifted up by one Floquet zone. 
    
    Corresponding experimental data (lower panel) shows quantized photon pumping. Fig.~\ref{fig:regimes}(a) corresponds to a semicircular protocol with $B=120$ MHz and $\Delta=101$ MHz.     
    
	\item \textit{Period-doubled trivial regime}. Adiabatic evolution over one external drive period $T$ maps a Floquet state onto the Floquet state of opposite type. Only after two periods, a total time of $2T$, does the state return to itself in the same Floquet zone. The quasi-energies satisfy,
	\begin{equation}
		\int_0^{4\pi}{\rm d}\theta_1 \,
		\frac{\partial \varepsilon(\theta_1)}{\partial \theta_1}
		=0,
	\end{equation}
	where $\varepsilon(\theta_1)$ denotes the quasi-energy branch followed adiabatically over the doubled cycle. During the first period, the cavity and external drive can exchange a non-quantized amount of energy, but the same amount is exchanged back in the second period, resulting in zero net energy transfer.
		
	Two qualitatively different examples of the quasi-energy spectrum in this regime are shown in the upper panels of Fig.~\ref{fig:regimes}(b.I) and (b,II). The spectrum features a single crossing per period between quasi-energy lines of opposite type. After two periods, a quasi-energy line reconnects back to itself.  
    
    The experimental data (lower panels) corresponding to Fig.~\ref{fig:regimes}(b.I) shows that the qubit state flips each period, but the cavity photon number remains the same. The system alternates between the two states $\approx \ket{0,g}$ and $\approx \ket{0,e}$ which differ by one polariton number. In contrast, the experimental data corresponding to Fig.~\ref{fig:regimes}(b.II) shows that the system alternates between the two states $\approx \ket{0,g}$ and $\approx \ket{1,e}$ that differ by two polariton numbers. Panel (b.I) corresponds to a semicircular protocol with $B=83$ MHz and $\Delta=101$ MHz; panel (b.II) corresponds to a protocol with maximum field amplitude $B=120$ MHz and $\Delta=101$ MHz, but the vertical stroke is offset by 30 MHz along $x$ (see inset). Note that in the upper panel of Fig.~\ref{fig:regimes}(b.II), the dotted circle indicates that the level crossing is not perfect, but has a gap that is too small to resolve. 
		
	\item \textit{Trivial regime}. Adiabatic evolution over one external drive period $T$ maps a Floquet state back to itself within the same Floquet zone. The quasi-energies satisfy,
	\begin{equation}
		\int_0^{2\pi}{\rm d}\theta_1 \,
		\frac{\partial \varepsilon_\pm(\theta_1)}{\partial \theta_1}
		=0.
	\end{equation}
	Consequently, there is no net energy transfer between the cavity and external drive.
		
	An example of the quasi-energy spectrum in the trivial regime is shown in Fig.~\ref{fig:regimes}(c). There are no level crossings, and each quasi-energy line reconnects to itself after one period. 
    
    The corresponding experimental data (lower panel) shows no photon pumping. Fig.~\ref{fig:regimes}(c) corresponds to an offset protocol with maximum field radius $B=120$ MHz, $\Delta=101$ MHz, but the vertical stroke is offset by 110 MHz along $x$. 
\end{enumerate}
In all upper panels of Fig.~\ref{fig:regimes}, the magnitude of the semi-classical cavity field is set to be $B_{\rm cav}=2g\sqrt{n}$, with $g=13$ MHz and $n=1$.

\paragraph{Semi-Classical Phase Diagram and Boundaries} Along the vertical segment of the protocol, quasi-energies are degenerate at specific points $B_z^\ell$ when the following condition is satisfied,
\begin{equation}
    \varepsilon_+(B_z^\ell) = \varepsilon_-(B_z^\ell) + \ell \Delta. 
    \label{eq:level_crossing_condition}
\end{equation}
The integer $\ell$ denotes the order of the crossing. In an $\ell$-th order crossing, the two degenerate quasi-energies are separated by $\ell$ Floquet zones. We note that, when the cavity is treated quantum mechanically, 
the integer $\ell$ corresponds to the difference in polariton numbers between 
the two degenerate states (more on this below).

The boundaries between the different topological regimes are determined by the number and order of quasi-energy crossings along the vertical segment $B_z\in(-B,B)$ of the protocol. Denote by $N_\ell$ the number of crossings of order $\ell$ satisfying $|B_z^\ell|<B$. Then, the quantity,
\begin{equation}
    N = \sum_\ell (N_\ell \bmod 2),
\end{equation}
gives a unique classification of the different topological regimes. In the pumping regime, $N=2$; in the period-doubled trivial regime $N=1$; in the trivial regime $N=0$. Note that the number of energy crossings of the same order can change by multiples of two within a topological regime. These correspond to processes where two crossings can annihilate without changing the winding of the quasi-energy spectrum. For example, this happens when two quasi-energy lines uncross as $\cloop \to \clines$, or vice-versa.

\begin{figure*}
    \centering
    \includegraphics{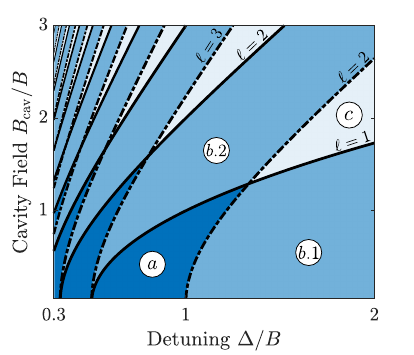}
    \caption{\textit{Semi-classical Phase Diagram of the Large--$\Delta$ Pump}. The color indicates the distinct topological regimes as a function of the two dimensionless ratios $B_{\rm cav}/B$ and $\Delta/B$ that define the semicircular protocol [Eqs.~\eqref{eq:H_pump_supp} and~\eqref{eq:B_fields_supp}]. The darkest shade denotes (a) the pumping regime, the second-darkest shade denotes (b) the period-doubled trivial regime, and the lightest shade denotes (c) the trivial regime. The labels (b.1) and (b.2) denote two distinct possibilities for the period-doubled trivial phase. In b.1(b.2), the system alternates between two different Floquet states that are separated by $\ell=1(2)$ Floquet zones. Examples of the quasi-energy spectra of each type are shown in the top panel of Fig.~\ref{fig:regimes}.}
    \label{fig:sc_phase_diagram}
\end{figure*}

The value of $N$ changes if (i) a new crossing $B_z^\ell$ enters the range $(-B,B)$, or (ii) an existing crossing leaves it. This occurs only when a crossing reaches one of the endpoints of the vertical segment, namely,
\begin{equation}
    B_z^\ell = +B
    \qquad (\textrm{top endpoint}),
    \qquad
    B_z^\ell = -B
    \qquad (\textrm{bottom endpoint}).
\end{equation}
At the top endpoint, the level crossing condition Eq.~\eqref{eq:level_crossing_condition} reads:
$\varepsilon_+(B)
=\varepsilon_-(B)+\ell\Delta$. It can be rewritten as,
\begin{equation}
	\frac{B_{\rm cav}}{B}
	=\sqrt{
	\left(\ell\frac{\Delta}{B}\right)^2
	-\left(1-\frac{\Delta}{B}\right)^2},
	\label{eq:top_condition}
\end{equation}
and only has real $B_{\rm cav}$ solutions for $\ell\geq 1$. 
Similarly, at the bottom endpoint, the condition
$\varepsilon_+(-B)
=\varepsilon_-(-B)+\ell\Delta$
can be written as,
\begin{equation}
	\frac{B_{\rm cav}}{B}
	=\sqrt{
	\left(\ell\frac{\Delta}{B}\right)^2
	-\left(1+\frac{\Delta}{B}\right)^2},
	\label{eq:bot_condition}
\end{equation}
and only has real $B_{\rm cav}$ solutions for $\ell \geq 2$.
The locus of points for which Eq.~\eqref{eq:top_condition} is satisfied is shown as solid black lines in Fig.~\ref{fig:sc_phase_diagram}. Each line corresponds to a different value of $\ell$. Similarly, solutions to Eq.~\eqref{eq:bot_condition} are shown as black dotted lines.

Finally, we note that the analysis above assumes that all exact level crossings are confined to the vertical segment of the protocol. This assumption is justified by symmetry: along the vertical segment, polariton number is conserved, so crossings between different polariton sectors are protected. Along the semicircular segment, by contrast, the transverse field is nonzero and couples different polariton number sectors, generically turning these crossings into avoided crossings. Exact crossings away from the vertical segment would therefore require additional fine tuning.

\paragraph{Connection to Quantum Phase Diagram} In this section, we now promote the cavity to a dynamical quantum degree of freedom and determine the pumping phase diagram shown in Fig.~\ref{fig:main_2}(b). We show that, starting from the vacuum, quantized pumping persists only up to a maximum photon number $n_{\rm max}$, for which we derive an analytic expression in Eq.~\eqref{eq:maximum_photon_number}. 

Within the semi-classical treatment of the cavity, if the system is initialized in the pumping regime, quantized energy transfer continues indefinitely in the adiabatic $\omega\to 0$ limit~\cite{Crowley:2019_classification, Vuina:2023ab}. The magnitude of the cavity field ${B}_{\rm cav}$ is a fixed external parameter that does not change in time. However, when the cavity is promoted to a dynamical quantum degree of freedom, the magnitude of the cavity field ${B}_{\rm cav}=2g\sqrt{n}$ itself increases in time as the cavity gains photons. By inspecting the phase diagram in Fig.~\ref{fig:sc_phase_diagram}, one expects that, starting from an empty cavity, there exists a maximum number of photons $n_{\rm max}$ that can be pumped into the cavity before $B_{\rm cav}$ grows large enough that the system exits the pumping regime~\cite{Nathan:2019aa}. 

The semi-classical phase diagram can be used to qualitatively understand the main features of the full quantum phase diagram in Fig.~\ref{fig:main_2}(b). The sharp cuts in pump performance occur at detunings $\Delta_\ell \approx B/\ell$ with $\ell=1,2,3,\dots$, corresponding to the boundaries between different semi-classical pumping regimes at $B_{\rm cav}=0$. However, in the quantum mechanical treatment of the cavity, the boundary curves receive small corrections controlled by $g/\Delta$, as we discuss below. 

The full problem with a quantum-mechanical cavity is also exactly solvable along the vertical segment of the protocol. Since the polariton number $n$ is conserved, eigenstates are labeled as $\ket{n,\pm}$, and the spectrum consists of two ladders,
\begin{align}
E_n^\pm &= \left(n-\frac{1}{2}\right)\Delta \pm\frac{1}{2}\sqrt{(B_z-\Delta)^2+4g^2 n}, \qquad n=1,2,3,\ldots
\label{eq:jc_energies}
\end{align}
where the two eigenstates $\ket{n,\pm}$ in a given polariton sector are superpositions of the bare states $\ket{n,g}$ and $\ket{n-1,e}$. The $n=0$ polariton sector contains the unique state $\ket{0,g}$, with energy $E_{0,g}=-B_z/2$. In the experiment, the system is initialized in the ground state $\ket{0,g}$. Note the similarity of the spectrum in Eq.~\eqref{eq:jc_energies} with the semi-classical Floquet quasi-energies Eq.~\eqref{eq:exact_quasi_energies}. They coincide with the replacement $B_{\rm cav} = 2g\sqrt{n}$.

In analogy to the semi-classical analysis, the topological classification is obtained by tracking level crossings of the form,
\begin{equation}
    E_n^+(B_z) = E_{n+\ell}^-(B_z).
    \label{eq:quantum_level_crossing}
\end{equation}
In the quantum analysis, the integer $\ell$ denotes the difference in polariton number between the two degenerate states. Again, this condition is almost identical to Eq.~\eqref{eq:level_crossing_condition}, except that the cavity field is no longer an independent parameter, but is itself fixed by the polariton number. Indeed, Eq.~\eqref{eq:quantum_level_crossing} can be written in terms of the semi-classical Floquet quasi-energies as $\varepsilon_+(B_z^\ell,B_{\rm cav}=2g\sqrt{n}) = \varepsilon_-(B_z^\ell,B_{\rm cav}=2g\sqrt{n+\ell}) + \ell \Delta$.

\begin{figure}
    \centering
    \includegraphics[scale=0.75]{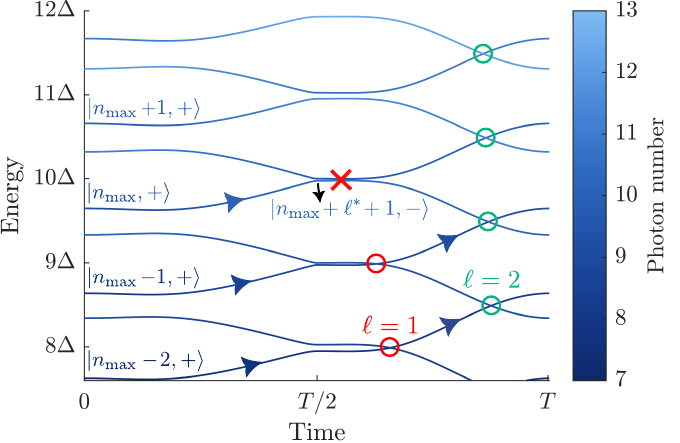}
    \caption{\textit{Energy Spectrum Above and Below the Maximum Photon Number.} Colored circles indicate exact level crossings. For states $\ket{n,+}$ with polariton number $n<n_{\rm max}$, adiabatic evolution along the semicircular segment transports the state one rung up the same $(+)$ eigenstate ladder, $\ket{n,+}\xrightarrow{ \text{circ}}\ket{n+1,+}$. The energy spectrum below $n_{\rm max}$ has the topology shown in Fig.~\ref{fig:regimes}(a), and the cavity absorbs one photon per drive period. For states $\ket{n,+}$ with polariton numbers $n\geq n_{\rm max}$, adiabatic evolution along the semicircular segment transports the state to the opposite $(-)$ eigenstate ladder, $\ket{n,+}\xrightarrow{ \text{circ}}\ket{n+\ell^\star+1,-}$, with $ \ell^\star = \left\lfloor{B}/{\Delta}+{g^2}/{\Delta^2}\right\rfloor$. The energy spectrum above $n_{\rm max}$ has the topology shown in Fig.~\ref{fig:regimes}(b.II), and there is no photon pumping. 
    As the polariton number increases, the boundary between the two regimes occurs at the point where the exact level crossing of order $\ell^\star$ disappears, as indicated by the red $\times$. The protocol parameters are $B=90~\mathrm{MHz}$ and $\Delta=82.5~\mathrm{MHz}$, corresponding to $\ell^\star=1$.}
    \label{fig:nMax}
\end{figure}

Like before, the quantum boundary curves are obtained by looking for level crossings at the two endpoints of the vertical segment, when $B_z=B$ or $B_z=-B$. At the top endpoint, the condition $E_n^+(B) = E^-_{n+\ell}(B)$ gives,
\begin{equation}
	\frac{B_{\rm cav}}{B}
	=\sqrt{
	\left[\ell\frac{\Delta}{B}\left(1-\frac{g^2}{\ell\Delta^2}\right)\right]^2
	-\left(1-\frac{\Delta}{B}\right)^2},
	\label{eq:top_condition_quantum}
\end{equation}
where $B_{\rm cav}=2g\sqrt{n}$. Similarly, the condition at the bottom endpoint $E_n^+(-B) = E^-_{n+\ell}(-B)$ gives,
\begin{equation}
	\frac{B_{\rm cav}}{B}
	=\sqrt{
	\left[\ell\frac{\Delta}{B}\left(1-\frac{g^2}{\ell\Delta^2}\right)\right]^2
	-\left(1+\frac{\Delta}{B}\right)^2}.
	\label{eq:bot_condition_quantum}
\end{equation}
Thus, in the limit $g/\Delta \ll 1$, these expressions reduce to the semi-classical boundary lines in Eqs.~\eqref{eq:top_condition} and~\eqref{eq:bot_condition}. This agreement can be understood from the effect of qubit backaction on the cavity. When the cavity is treated as a dynamical variable, its coupling to the qubit renormalizes its effective frequency, as discussed in the Supplementary Materials of Ref.~\cite{Long2022b}. This frequency shift is controlled by the small parameter $g/\Delta$ and becomes negligible for $g/\Delta\ll1$. In this regime, the cavity evolves independently at its bare frequency $\Delta$, justifying the semi-classical description.

Observe from the semi-classical phase diagram in Fig.~\ref{fig:sc_phase_diagram} that, by increasing $B_{\rm cav}$ while holding $\Delta$ fixed, the exit from the pumping regime always happens by crossing one of the solid black curves. This corresponds to the appearance of a level crossing of order $\ell^\star$ at the top endpoint $B_z=B$, with,
\begin{equation}
    \ell^\star = \left\lfloor{B}/{\Delta}+{g^2}/{\Delta^2}\right\rfloor.
    \label{eq:ell_value}
\end{equation}
The quantity $\ell^\star+1$ corresponds to the number of perfect level crossings between the ground state $\ket{0,g}$ and other higher polariton number states $\ket{n,-}$ throughout the vertical segment of the protocol. It is also the integer for which the detuning satisfies $\Delta_{\ell^\star+1}<\Delta<\Delta_{\ell^\star}$, with $\Delta_\ell$, defined in Eq.~\eqref{eq:critical_detunings} of the main text. 

The maximum photon number is then the largest value of $n$ reached before the system crosses the corresponding $\ell^\star$ boundary line. Thus, $n_{\rm max}=\left\lfloor(B_{\rm cav}(\ell^\star)/(2g))^2\right\rfloor$, or equivalently,
\begin{equation}
    n_{\rm max} = \left\lfloor \frac{B^2}{4g^2} \left\{ \left[
    \ell^\star\frac{\Delta}{B} \left(1-\frac{g^2}{\ell^\star\Delta^2}
    \right)\right]^2 - \left(1-\frac{\Delta}{B}\right)^2\right\}\right\rfloor .
    \label{eq:maximum_photon_number}
\end{equation}
Plotting Eq.~\eqref{eq:maximum_photon_number} as a function of $\Delta$ and $B$, for $g=13$ MHz, gives the phase diagram shown in Fig.~\ref{fig:main_2}(b) of the main text. Moreover, since $\ell^\star\leq B/\Delta+g^2/\Delta^2$, one has
\begin{equation}
    \ell^\star\frac{\Delta}{B}\left(1-\frac{g^2}{\ell^\star\Delta^2}\right)\leq1,
\end{equation}
which implies the upper bound,
\begin{equation}
    n_{\rm max}\leq \frac{B^2}{4g^2}.
    \label{eq:n_max}
\end{equation}
Experimental data for the phase diagram, including at the $\Delta=\Delta_2$ resonance and the trivial regime beyond $\Delta >\Delta_1$, as well as all three scenarios discussed in Fig.~\ref{fig:main_2}, are shown in Fig.~\ref{fig:detuning_sweep_sup}. Near $\Delta\approx \Delta_2$, the pump exhibits small and irregular energy transfer to the cavity, but no systematic pumping, as shown in Fig.~\ref{fig:detuning_sweep_sup}(d). In the trivial regime beyond $\Delta > \Delta_1$, one has $n(mT) = 0$, as shown in Fig.~\ref{fig:detuning_sweep_sup}(e).

\paragraph{Breakdown of pumping} For $n<n_{\rm max}$, adiabatic evolution along the semicircular segment of the protocol results in,
\begin{equation}
    \ket{n,+} \xrightarrow{\text{circ}}  \ket{n+1,+}.
\end{equation}
Note that for $(B_z-\Delta)^2\gg g^2n$ and $B_z<\Delta$, we have $\ket{n,+}\approx\ket{n,g}$. Likewise, for $(B_z-\Delta)^2\gg g^2n$ and $B_z>\Delta$, we have $\ket{n+1,+}\approx\ket{n,e}$. Thus, evolution along the semicircular segment approximately conserves the photon number while flipping the state of the qubit $\ket{n,g} \xrightarrow{ \text{circ}}  \ket{n,e}$.

Once the system reaches $n=n_{\rm max}$, adiabatic evolution along the semicircular segment instead results in,
\begin{equation}
    \ket{n_{\rm max},+} \xrightarrow{ \text{circ}} \ket{n_{\rm max}+\ell^\star+1,-}.
\end{equation}
That is, the system is adiabatically transported to the state with $\ell^\star+1$ more polaritons of the opposite (-) branch. This is a depumping state, and the cavity begins to lose photons on subsequent cycles. Under the same approximation as above, $\ket{n_{\rm max},+}\approx \ket{n_{\rm max},g}$ and $\ket{n_{\rm max}+\ell^\star+1,-}\approx\ket{n_{\rm max}+\ell^\star+1,e}$, we see that the cavity approximately gains $\ell^\star+1$ photons before depumping begins. 

At the finite frequency $\omega$ used in the experiment, the system is not able to adiabatically transition to a depumping state upon reaching $n_{\rm max}$. Instead, Fig.~\ref{fig:main_2}(e) shows a sudden loss of qubit contrast, which is indicative of non-adiabatic processes that result in a mixed qubit state. Fig.~\ref{fig:pump_breakdown} shows numerical data that supports this.

The spectrum immediately above and below $n_{\rm max}$ is shown in Fig.~\ref{fig:nMax}. For polariton numbers $n<n_{\rm max}$, the energy spectrum has the topology of the pumping regime in Fig.~\ref{fig:regimes}(a); the cavity absorbs one photon per period $T$. For polariton numbers $n> n_{\rm max}$, the spectrum has the topology of the period-doubled trivial regime in Fig.~\ref{fig:regimes}(b.II); there is no quantized photon pumping. The protocol parameters for the example spectrum shown are $B=90$ MHz and $\Delta=82.5$ MHz, corresponding to $n_{\rm max}= 9$,  $\ell^\star = 1$.

\subsection{Chern Numbers}
\label{sec:chern_numbers}
\begin{figure*}
    \centering
    \includegraphics[width=\linewidth]{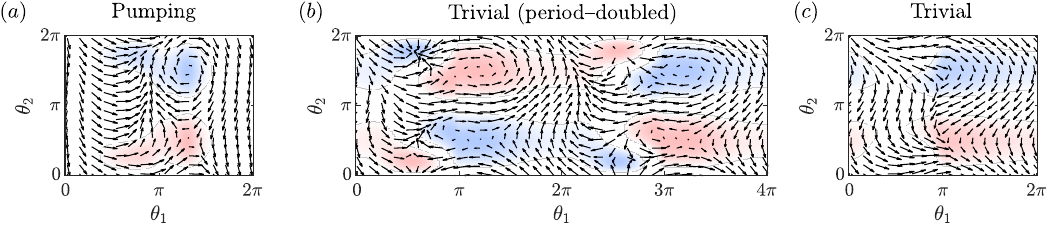}
    \caption{\textit{Chern Numbers.} Spin textures of the dressed adiabatic state $\ket{\widetilde{\psi}(\theta_1,\theta_2)}$ followed by the qubit in each of the three distinct topological regimes. Black arrows show the projection of the qubit state onto the $(x,z)$ plane, with components $(\langle\sigma_x\rangle,\langle\sigma_z\rangle)$, while color encodes $\langle\sigma_y\rangle$ from deep blue $(-1)$ to deep red $(+1)$. (a) In the pumping regime, the adiabatic state defines a texture with nonzero Chern number in the $(\theta_1,\theta_2)\in[0,2\pi)^2$ torus. (b) In the period-doubled trivial regime, the adiabatic state defines a texture with zero Chern number in the extended torus $(\theta_1,\theta_2)\in[0,4\pi)\times[0,2\pi)$. (c) In the trivial regime, the adiabatic state defines a texture with zero Chern number in the $(\theta_1,\theta_2)\in[0,2\pi)^2$ torus.}
    \label{fig:supp_chern_numbers}
\end{figure*}

Topological photon pumping can alternatively be understood in terms of the Chern number of the dressed adiabatic states $\ket{\widetilde{\psi}(\theta_1,\theta_2)}$ followed by the qubit. Recall that the dressed adiabatic states $\ket{\widetilde{\psi}(\theta_1,\theta_2)}$ are defined from the solution to the Floquet problem over the fast cavity angle $\theta_2$ at fixed slow drive angle $\theta_1$, as in Eq.~\eqref{eq:Floquet_solutions}. Figure~\ref{fig:supp_chern_numbers} shows the spin textures of the dressed adiabatic states, as a function of the two angles $(\theta_1,\theta_2)$, and in each of the three distinct topological regimes. The black arrows show the projection of the qubit state onto the $(x,z)$ plane, and have components $(\langle\sigma_x\rangle,\langle\sigma_z\rangle)$. The remaining component $\langle\sigma_y\rangle$ is indicated by color, with deep red corresponding to $\langle\sigma_y\rangle=+1$ and deep blue to $\langle\sigma_y\rangle=-1$.

In the pumping regime [Fig.~\ref{fig:supp_chern_numbers}(a)], the dressed adiabatic state is periodic over one drive period $\theta_1\to\theta_1+2\pi$ and therefore defines a texture on the torus $(\theta_1,\theta_2)\in[0,2\pi)^2$. The corresponding Chern number is nonzero, leading to quantized energy pumping between the cavity and external drive.

In the period-doubled trivial regime [Fig.~\ref{fig:supp_chern_numbers}(b)], the dressed adiabatic state is \emph{not} periodic over one drive period $\theta_1\to\theta_1+2\pi$, but over two: $\theta_1\to\theta_1+4\pi$. The adiabatic state therefore defines a texture on the extended torus $(\theta_1,\theta_2)\in[0,4\pi)\times[0,2\pi)$. While the cavity and external drive may exchange a non-quantized amount of energy over a single drive period, the corresponding Chern number on the extended torus is zero, resulting in zero net energy exchange over two drive periods.

Finally, in the trivial regime [Fig.~\ref{fig:supp_chern_numbers}(c)], the dressed adiabatic state is periodic over one drive period $\theta_1\to\theta_1+2\pi$, and defines a texture on the torus $(\theta_1,\theta_2)\in[0,2\pi)^2$. The Chern number is zero, resulting in zero net energy pumping even under a single drive period.

\section{Processes that degrade or destroy photon pumping}
\label{sec: pump breakdown}
\begin{figure*}
\centering
    \includegraphics{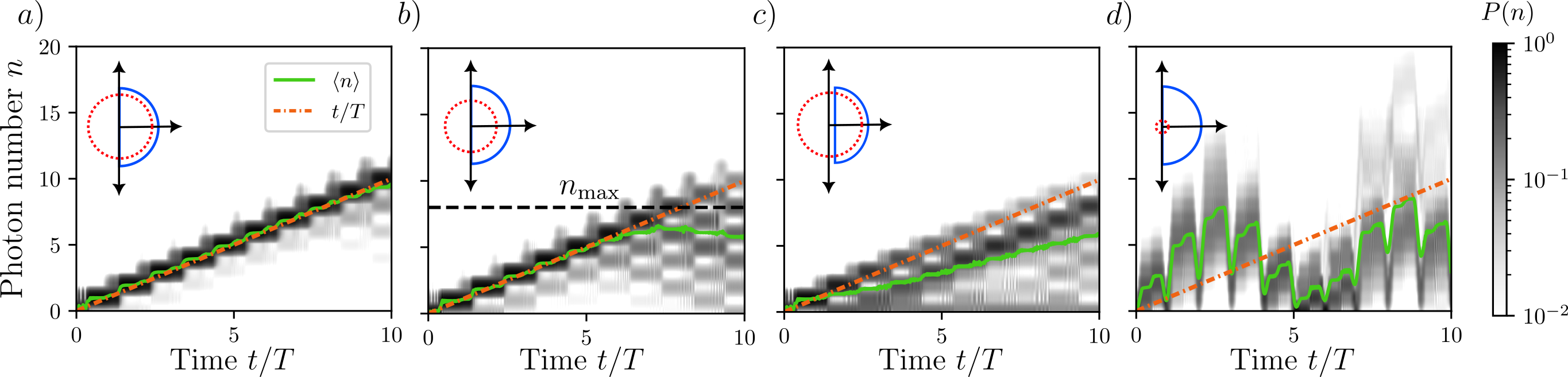}
    \caption{\textit{Processes that Degrade or Destroy Photon Pumping}. Numerical simulations of the cavity photon number distribution vs time in four distinct regimes of operation: (a) ideal pump operation at $\Delta=100~\mathrm{MHz}$; (b) pump operation limited by the theoretical upper bound $n_{\rm max}$ at $\Delta=85~\mathrm{MHz}$; (c) pump operation degraded by a small $10~\mathrm{MHz}$ offset from the ideal semicircular protocol at $\Delta=100~\mathrm{MHz}$; and (d) pump operation hindered by strong backaction on the cavity at small detuning $\Delta=6~\mathrm{MHz}$. All simulations are dissipationless, with drive frequency $\omega=4~\mathrm{MHz}$ and maximum field amplitude $B=120~\mathrm{MHz}$. The color denotes the instantaneous Fock-state population $P(n)=|\braket{n|\Psi(t)}|^2$ on a logarithmic scale. The green and red lines denote the average photon number in the cavity at each $t$, and the ideal photon number expected for the constant quantized pumping rate of $P_q$.}
    \label{fig:pump_breakdown}
\end{figure*}

Photon pumping can break down in several ways. Fig.~\ref{fig:pump_breakdown} shows numerical simulations of the intracavity photon-number populations over ten drive periods in four distinct settings. The color map denotes the instantaneous Fock-state population $P(n)=|\braket{n|\Psi(t)}|^2$ on a logarithmic scale. All simulations are performed without cavity dissipation. In all panels, the drive frequency is $\omega=4~\mathrm{MHz}$, and the maximum field amplitude is $B=120~\mathrm{MHz}$.

We now discuss each panel.
\begin{enumerate}[label=(\alph*)]
    \item shows ideal pump operation at $\Delta=100~\mathrm{MHz}$. The mean cavity photon number closely follows the ideal quantized form $\langle n(t)\rangle=t/T$.

    \item shows pumping limited by the theoretical upper bound $n_{\rm max}$ at $\Delta=85~\mathrm{MHz}$. In the adiabatic limit $\omega\to0$, upon reaching the upper bound $n_{\rm max}$, the system would transition into a depumping state, after which the cavity would begin to lose photons on subsequent cycles. However, at the finite drive frequency used here, we instead observe a gradual loss of pumping: part of the wave function enters a depumping state, while another part diabatically continues to pump for additional cycles. The qubit state becomes mixed, which explains the sudden loss of contrast observed in the experimental data in Fig.~\ref{fig:main_2}(e).

    \item shows pumping degraded by a $10~\mathrm{MHz}$ offset of the vertical stroke of the semi-circle at $\Delta=100~\mathrm{MHz}$. Photon pumping is not completely lost, but its quality is degraded, visible as a deviation of $\langle n(t)\rangle$ from the ideal slope. The offset opens a small gap at both the $\ell=1$ and $\ell=2$ level crossings shown in Fig.~\ref{fig:regimes}(a), which would otherwise be protected by polariton number conservation. 

    \item shows pumping destroyed by strong qubit-to-cavity backaction at small detuning $\Delta=6~\mathrm{MHz}$. The mean photon number undergoes large excursions within each drive period, and there is no signature of quantized pumping. The stroboscopic photon number $n(mT)$ also exhibits large period-to-period variations, consistent with the experimental data in Fig.~\ref{fig:main_2}(f) of the main text. 
\end{enumerate}

\begin{figure*}[t]
    \centering
    \includegraphics[width=\linewidth]{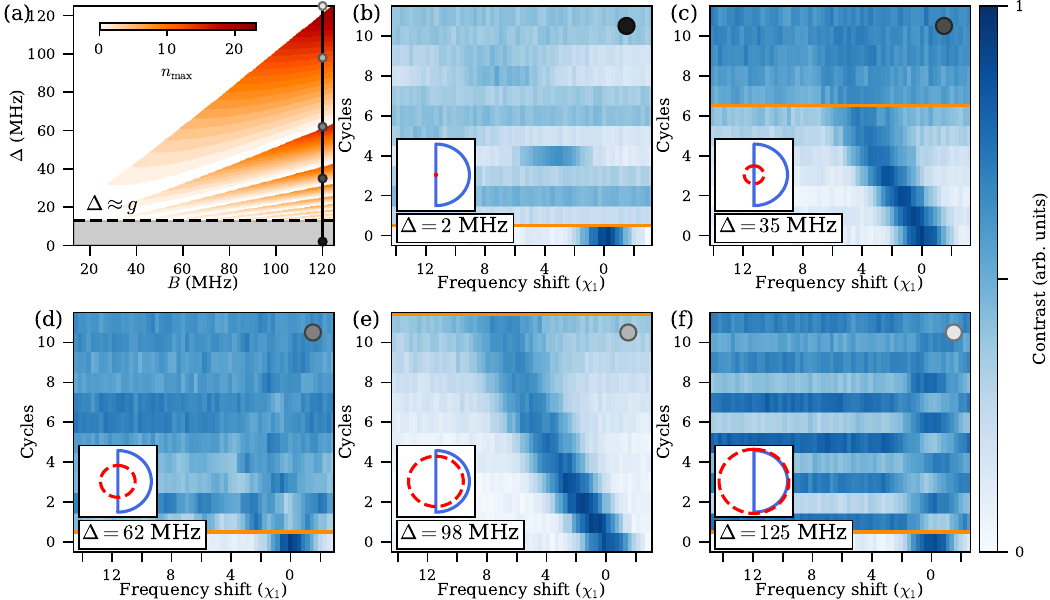}
    \caption{\textit{Extended Data of Photon Pump Phase Diagram.} (a) The analytically computed maximum photon number $n_{\max}$ (color) as a function of $B$ and $\Delta$ for $g = 13$~MHz, neglecting cavity loss. Panels (b-f) show spectroscopy of the AC Stark shift on the qubit versus number of drive cycles along the black line $B = 120$~MHz, corresponding to markers in (a). Panels (b-c) and (e) show additional data taken in the small $\Delta$ back-action regime, $n_{\mathrm{max}}$-limited regime, and $\kappa$-limited regime, respectively. In addition to the three scenarios, panels (d) and (f) show data with detuning $\approx\Delta_{2}$ and $>\Delta_{1}$ both discussed in Eq.~\ref{eq:critical_detunings}, respectively. Drive protocol and detuning for each panel shown in inset. The orange line indicates the cutoff cycle used to extract $n_{\mathrm{final}}$ (see Sec.~\ref{sec:experimental_realization} for details). 
    }
    \label{fig:detuning_sweep_sup}
\end{figure*}

\section{Quantized Pumping}
\label{sec:quantized pumping}
\begin{figure}[t]
\centering
    \includegraphics[width=\linewidth]{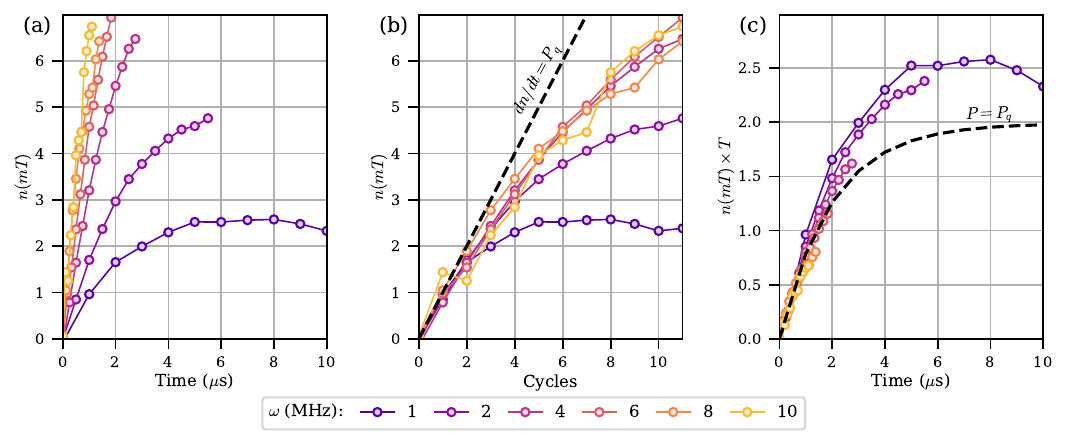}
    \caption{\textit{Quantized Time Dynamics of the Photon Pump.} Photon number $n(mT)$ for $\omega$ between $1$~MHz and $10$~MHz at $B=120$~MHz and $\Delta=101$~MHz (a) versus time, (b) versus number of drive cycles. Over short times, the observed photon number closely follows the ideal pump rate $dn/dt = P_q=1/T$, marked with the dashed line in (b), but saturates at long times due to cavity loss and imperfect adiabatic following. (c) Period-normalized photon number $n(mT)\times T$ versus time. At low $\omega$, photon addition at a fixed point during the drive protocol results in deviations from $n(mT)=(P_q/\kappa)\times (1-e^{-\kappa mT})$ (dashed line). At intermediate values of $\omega$, deviations are small and observed pump rates closely match the ideal quantized rate.}
    \label{fig:omega_sweep_sup}
\end{figure}
\begin{figure*}[ht]
    \centering
    \includegraphics[width=\linewidth]{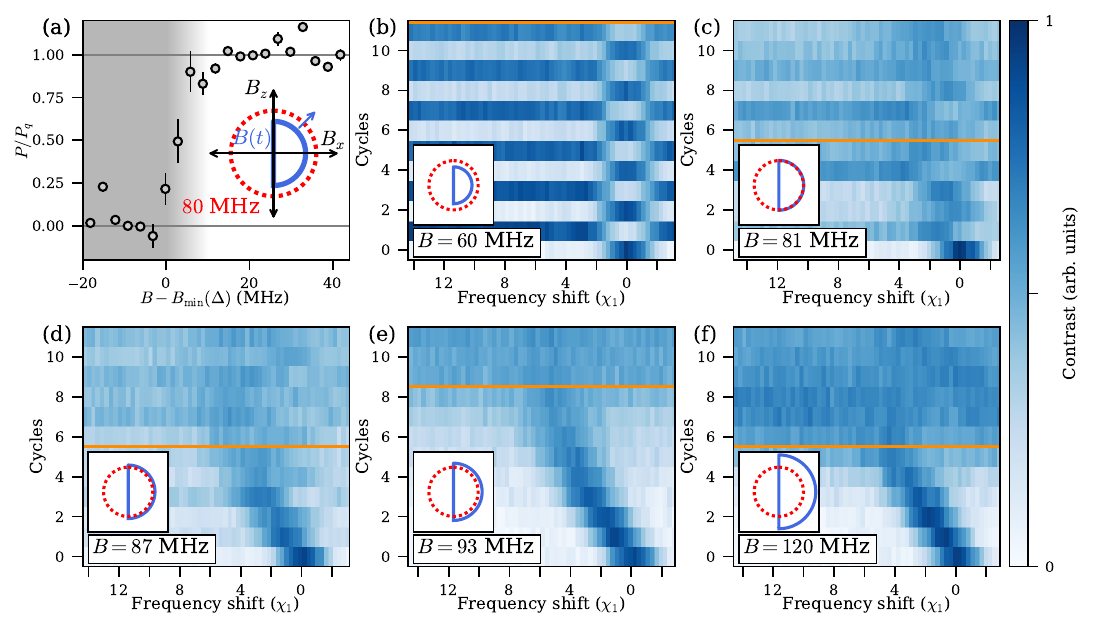}
    \caption{\textit{Quantized Pump Rate at $\Delta = 80\ \mathrm{MHz}$.} (a) Photon pump rate $P/P_q$ extracted from the first five cycles versus $B$ for $\Delta = 80$~MHz and $\omega=5$~MHz. Error bars are determined from the fit to $n(mT) = (P/\kappa) \times (1 - e^{-\kappa mT})$.
    (b-f) Spectroscopy of the AC Stark shift on the qubit versus number of drive cycles. Drive protocol and field size for each panel are shown in the inset. The orange line indicates the cutoff after which the qubit frequency cannot be resolved (see Sec.~\ref{sec:experimental_realization} for details).
    (b) $B<B_{\mathrm{min}}(\Delta)$, where the qubit state flips every cycle and the pump exhibits no energy transfer. 
    (c) $B\approx B_{\mathrm{min}}(\Delta)$, where the pump exhibits small and inconsistent energy transfer.
    (d-f) $B> B_{\mathrm{min}}(\Delta)$, where the pump lies in the region between $\ell^\star =1$ and $\ell^\star =2$ and exhibits energy transfer with varying $n_{\mathrm{max}}$ set by the exact configuration of the drive. While $n_{\mathrm{final}}$ differs between configurations, the energy transfer rate is consistent throughout this regime at early times.}
    \label{fig:b_sweep_sup}
\end{figure*}

In this section, we present additional data that demonstrate the quantized pump rate of the photon pump in the topological regime. Specifically, we examine the behavior of the pump at rotation rates $\omega$ between $1$~MHz and $10$~MHz at fixed $\Delta=101$~MHz and $B = 120~\mathrm{MHz}$. The measured intracavity photon number versus time for each pump frequency is shown in Fig.~\ref{fig:omega_sweep_sup}(a).
An ideal pump would display a universal pump rate $P_q = 1/T = 2\pi/\omega$, independent of $\omega$. 
To demonstrate this behavior, Fig.~\ref{fig:omega_sweep_sup}(b) shows the number of cavity photons versus the number of completed cycles of the drive. At early times and for sufficient pump rates, the observed $dn/dt$ lies close to this ideal line, but never saturate the ideal pump rate due to the presence of cavity loss and imperfect adiabatic following during the pump cycle. At long times the cavity photon number saturates because of the competing influences of the drive protocol and cavity loss.

The effect of weak cavity loss on photons by a pump with pump rate $P$ can be captured by a simple model: 
\begin{equation}
n(mT) \approx \frac{P}{\kappa} \left( 1-e^{-\kappa m T} \right),
\label{eq_sup:pumpingrate}
\end{equation}
where $T$ is the period of the drive protocol, and $\kappa$ is the cavity loss rate. As $PT$ is a constant, $n(mT) \times T$ versus $mT$ should collapse onto a single curve for all $\omega$, and hence Eq.~\ref{eq_sup:pumpingrate} can be used to determine $P$ even in the presence of photon loss. 
Fig.~\ref{fig:omega_sweep_sup}(c) shows $n(mT) \times T$ for all $\omega$. 
We find that our photon pump closely matches this model and exhibits a pump rate close to $P_q = 1/T$ at $\omega \gtrsim 4$~MHz
The deviation at small $\omega$, is expected, since Eq.~\ref{eq_sup:pumpingrate}  assumes a continuous accumulation of cavity photons, whereas in actuality the photon pump adds a photon at a fixed point during the drive protocol.
At higher values of $\omega$, imperfect adiabatic following reduces the effective pump rate, causing the measured photon population to fall slightly below the ideal rate.

To extract the values of $P$ reported in Fig.~\ref{fig:main_3} in the main section of the manuscript, we fit data for $\omega=5$~MHz over the first five drive periods, or until the photon number is not resolvable if this occurs earlier. Details of the photon number cutoff criteria can be found in Sec.~\ref{sec:experimental_realization}. We find that the fit to Eq.~\ref{eq_sup:pumpingrate} is good and produces pump rates $P \approx 0$ in the trivial regime and $P \approx P_q$ in the topological regime across a range of $B$ and $\Delta$ values, as discussed in the main section of the manuscript. Fig.~\ref{fig:b_sweep_sup} shows sample spectroscopy traces at varying $B$ and fixed $\Delta=80$~MHz used in this analysis.

\section{Experimental Realization}
\label{sec:experimental_realization}
\subsection{Device and Wiring}
The device consists of a tunable transmon qubit capacitively coupled to a storage cavity and a readout resonator \cite{Blais_review,quantum_engineer_guide} with the system parameters listed in Table.~\ref{tab:qubit_params}. The device was fabricated on a sapphire substrate with a superconducting tantalum (Ta) film. The Josephson junctions consisting of Al/AlO$_x$/Al tunnel barriers were fabricated using electron-beam lithography followed by double-angle evaporation. Other chip features were patterned using standard photolithography and wet etching. A detailed description of the device design and fabrication process can be found in Ref.~\cite{toolkitmartin}.

The room-temperature experimental setup and fridge wiring can also be found in Ref.~\cite{toolkitmartin}, with the 5 MHz low-pass filter on the flux-bias line replaced by an 80 MHz low-pass filter to reduce distortion and increase the maximum achievable modulation rate. 
\begin{table*}[t]
    \centering
    \begin{tabular}{|c|c|c|}
    \hline
    Parameter & Symbol & Value \\
       \hline \hline
       Qubit g-e frequency & $\omega_q$/(2$\pi$)  & 3.9-7.4 GHz\\
       \hline
       Qubit anharmonicity & $\alpha$/(2$\pi$)    & 240 MHz \\
       \hline
       Qubit-boost cavity coupling & $g_m$/(2$\pi$)       & 13 MHz\\
       \hline
       Qubit-readout cavity coupling & $g_r$/(2$\pi$)       & 90 MHz\\
       \hline 
       Qubit decay rate & $\Gamma_q$/(2$\pi$) & 13.9 kHz \\
       \hline \hline
       Readout cavity frequency & $\omega_r$/(2$\pi$)  & 7.492 GHz\\
       \hline
       Readout cavity linewidth & $\kappa_r$/(2$\pi$)  & 350 kHz\\
       \hline \hline 
       Storage cavity frequency & $\omega_m$/(2$\pi$)  & 5.04 GHz\\
       \hline
       Storage cavity linewidth & $\kappa_m$/(2$\pi$)  & 84 kHz\\
       \hline
    \end{tabular}
    \caption{\textit{Device Parameters}.}
    \label{tab:qubit_params}
\end{table*}
\subsection{Hamiltonian Implementation}
\label{sec:Hamiltonian Implementation}
In this section, we provide additional details on the implementation of the effective Hamiltonian used throughout the manuscript. The experiment is realized in a superconducting circuit platform consisting of a transmon qubit capacitively coupled to two coplanar waveguide cavities. We use the two lowest-energy states of the transmon, denoted by $\ket{g}$ and $\ket{e}$, to define the qubit, with $\omega_q$ corresponding to the energy difference between these two states~\cite{Blais_review,quantum_engineer_guide}. One cavity (storage) is used to store pumped photons. The other cavity (readout) serves as qubit state readout and provides microwave control to the qubit. Neglecting higher transmon levels and the far-detuned readout resonator, the system is described by the driven Jaynes--Cummings Hamiltonian
\begin{equation}
H^{\mathrm{bare}}=\omega_c a^\dagger a+\frac{\omega_q}{2}\sigma_z+g(a^\dagger\sigma^-+a\sigma^+)+\frac{\Omega}{2}\cos(\omega_d t)\sigma_x ,
	\label{eq_sup:bz}
\end{equation}
where $\omega_c$ and $\omega_q$ are the cavity and qubit frequencies, $g$ is the coupling strength between the qubit and the storage cavity, and $\Omega$ and $\omega_d$ are the Rabi rate (amplitude) and frequency of the microwave drive. 

The pumping protocol requires periodic modulation of the qubit with an effective Hamiltonian
\begin{equation}
H^{\rm{eff}} = \vec{B}(t)\cdot \vec{\sigma},
\end{equation}
using a magnetic field 
\begin{equation}
	\vec{B}(t) = \left[ B\max\left[\sin(\omega t),\,0\right],0,-B\cos(\omega t)\right]
	\label{eq_sup:semi_circle_protocol}
\end{equation}
that traces out a semi-circular trajectory in the $xz$-plane with amplitude $B$ and rotation rate $\omega$.
Because a transmon qubit does not possess a physical spin degree of freedom, we synthesize an \emph{effective} magnetic field in a rotating frame, following the method described in Refs.~\cite{toolkitmartin, Long2022b}. The $z$-component of the field is realized by flux modulation of the qubit frequency~\cite{Blais_review, quantum_engineer_guide}, while the $x$-component is generated by amplitude modulation of a resonant microwave drive. We use an external DC magnet to bias the qubit to a desired frequency $\omega_q$. An on chip flux-bias line is used to modulate the qubit about $\omega_q$ to produce:
\begin{equation}
	H^{\rm{bare}}_z(t) = \frac{(\omega_q-B\cos(\omega t))}{2}\sigma_z    \rightarrow H^{\rm{eff}}_z(t) = \frac{-B}{2}\cos(\omega t) \sigma_z,
	\label{eq_sup:bz}
\end{equation}
where $B$ and $\omega$ are the modulation strength and rate of the flux-bias line. 

The $x$-component of the effective magnetic field is realized by driving the qubit on resonance with a microwave drive (X-drive) through the readout cavity. In the rotating frame of the applied drive $\omega_d=\omega_q$, the system can be described as $H^{\rm{eff}}_x = \frac{\Omega}{2}\sigma_x$, where $\Omega$ is the Rabi rate of the microwave drive, which is linearly proportional to the amplitude of the applied drive. By applying a half-wave-rectified sinusoidal modulation to the Rabi rate, we achieve the desired transverse field: 
\begin{equation}
	H^{\rm{eff}}_x(t) = \frac{B\max\left[\sin(\omega t),\,0\right]}{2}\sigma_x .
	\label{eq_sup:bx}
\end{equation}

When characterizing the possible winding configurations of the Floquet states (Sec.~\ref{sec:topological_classification}) or examining mechanisms that cause pumping to break down (Sec.~\ref{sec: pump breakdown}), we allow one more degree of freedom in the transverse field and incorporate 
a variable offset $B_{\rm{offset}}$ in the transverse direction, while keeping maximum field amplitude constant at $B$. Specifically, the effective transverse field has the following form:
\begin{equation}
	H^{\rm{eff}}_x(t) = \frac{\max\left[(B-B_{\mathrm{offset}})\sin(\omega t)+B_{\mathrm{offset}},\,B_{\mathrm{offset}}\right]}{2}\sigma_x .
	\label{eq_sup:bx_offset}
\end{equation}

Combining both control fields, setting $\omega_q = \omega_d$, and going to the rotating frame of the applied drive using $U(t) = \exp[-i\omega_d t(a^\dagger a + \sigma_z/2)]$ yields the effective Hamiltonian 
\begin{equation}
	H_{\mathrm{pump}}(t) = \Delta\, a^\dagger a + g(a^\dagger \sigma^- + a \sigma^+) + \frac{1}{2}\vec{B}(t)\cdot\vec{\sigma},
	\label{eq_sup:JCH_rot}
\end{equation}
where $\Delta = \omega_c - \omega_d$, as promised by Eq.~\ref{eq:JCH_rot}~\cite{Long2022b,toolkitmartin}.

Accurate extraction of the photon number requires a fixed bare qubit frequency during qubit spectroscopy. Prior to each measurement, an active servo on the DC magnet stabilizes the qubit frequency at a designated operating point $\omega_{q_0}$, chosen to optimize qubit spectroscopy. In general, $\omega_{q_0}\neq\omega_d$. To ensure that the mean qubit frequency $\omega_q$ coincides with the drive frequency $\omega_d$, in the protocol sequence, a constant offset is applied through the flux-bias line in addition to the sinusoidal modulation described above. Further details regarding field calibration and characterization can be found in Refs.~\cite{dissipation_paper_PRX,toolkitmartin}.

\subsection{Measurement Protocol}
\begin{figure*}[t]
    \centering
    \includegraphics[width=\linewidth]{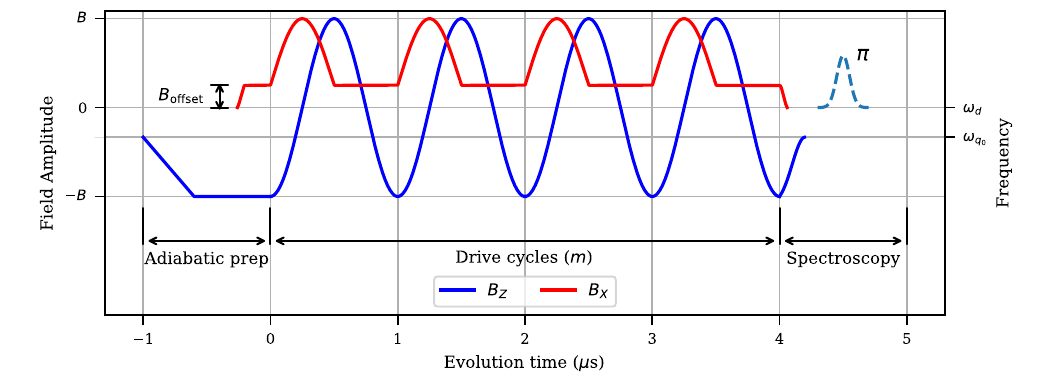}
    \caption{ \textit{Protocol schedule.} 
    The experiment protocol includes three parts: adiabatic initialization, evolving $H_{\mathrm{pump}}$ for $m$ cycles, and qubit ramp-out followed by photon-number measurement. The qubit is initialized in the ground state with transition frequency $\omega_{q_0}$ optimized for photon-number measurement.
    A slow ramp in $B_z$, while keeping $B_x = 0$, followed by a slow ramp in $B_x$ brings $\ket{0,g}$ to an eigenstate of $H_{\mathrm{pump}}(t=0)$, independent of $\vec{B}(t=0)$~\cite{toolkitmartin}.
    To detect the intracavity photon number, the qubit is adiabatically brought back to $\omega_{q_0}$, mapping the instantaneous eigenstate of $H_{\mathrm{pump}}(t)$ to the $z$-axis of Bloch sphere. Qubit spectroscopy is performed using a $\pi$-pulse with variable frequency.}
    \label{fig:schedule}
\end{figure*}
In this section, we describe the experimental protocol used throughout the manuscript. Each measurement consists of three stages: adiabatic initialization, evolution under the effective magnetic field $\vec{B}(t)$ for a variable number of drive cycles $m$, and qubit ramp-out followed by photon-number measurement, as shown in Fig.~\ref{fig:schedule}. 

The goal of adiabatic initialization is to prepare the system into a pumping state of the instantaneous spectrum of $H_{\mathrm{pump}}(t)$. This is accomplished by starting with the qubit in the $\ket{g}$ state with the effective field off and slowly ramping to the desired $B_z$, while $B_x = 0$. We then slowly ramp up $B_x$. This brings the vacuum state $\ket{0,g}$ to an eigenstate of $H_{\mathrm{pump}}(t=0)$, independent of $\vec{B}(t=0)$~\cite{toolkitmartin}.

After evolving the system under $\vec{B}(t)$ for $m$ cycles, we probe the cavity state using the same qubit. 
After completing the desired pump cycles, we return the qubit adiabatically to $\omega_{q_0}$ using a Gaussian flux ramp designed to reduce potential distortion from flux-bias line filtering, while also slowly turning $B_x$ to zero. This maps the instantaneous eigenstates of $H_{\mathrm{pump}}(t)$ to the $z$-axis of Bloch sphere.
We then perform two-tone spectroscopy \cite{Blais_review,quantum_engineer_guide} to determine the qubit frequency. A microwave pulse ($\pi$ pulse) with variable frequency $\omega_{\mathrm{spec}}$ is sent through the readout cavity, which flips the qubit state if $\omega_{\mathrm{spec}}$ equal to the AC Stark shifted qubit frequency. The resulting peak/dip in spectroscopy contrast indicates the shifted qubit frequency.
The spectroscopy $\pi$ pulse is generated using the same microwave source as the X-drive, whose carrier frequency is fixed at $\omega_d$. The pulse is synthesized using single-sideband (SSB) modulation at an intermediate frequency $\omega_{\mathrm{IF}}=\omega_{\mathrm{spec}}-\omega_d$, which produces a single sideband at the desired spectroscopy frequency $\omega_{\mathrm{spec}}$. By varying $\omega_{\mathrm{IF}}$, the effective frequency of the spectroscopy pulse can be swept without changing the carrier frequency of the microwave source. 

$\omega_{q_0}$ is set to 4.90 GHz for all measurements except for single-photon-number-resolved measurements shown in Fig.~\ref{fig:main_4}. For single-photon-number-resolved measurements, $\omega_{q_0}$ is increased to 5.00 GHz to increase the AC Stark shift per photon. Each frequency point is averaged 2,000 times.

\subsection{Photon-Number Calibration and Extraction}
\label{sec:n extraction}

We probe the intracavity photon population by measuring the AC Stark shift on the qubit. In the dispersive regime, where the qubit-cavity detuning is much larger than the coupling strength, $|\Delta_{\mathrm{qc}}|=|\omega_{q_0}-\omega_c|\gg g$, the Jaynes-Cummings Hamiltonian has the dispersive form:
\begin{equation}
H_{\mathrm{disp}} \approx  \omega_ca^\dagger a+\frac{1}{2}\omega_q\sigma_z + \frac{1}{2}\chi_1 a^\dagger a\sigma_z, \label{eqn:JCdisp}
\end{equation}
where $\chi_1 = -\frac{2g^2\alpha}{\Delta_{\mathrm{qc}}(\Delta_{\mathrm{qc}}-\alpha)}$ is a 1-photon shift for a transmon qubit with anharmonicity $\alpha$ \cite{Blais_review, quantum_engineer_guide}. As a result, $n$ photons in the cavity produce a linear AC Stark shift on the qubit $\chi_n =  n\chi_1$. 
For almost all data in the manuscript, we use $\omega_{q_0} = 4.9$~GHz and $\Delta_{\mathrm{qc}} = -140$~MHz, where the single-photon AC Stark shift $\chi_1 = -1.74$ MHz, determined from numerical calculation with QuTip \cite{qutip} and the chip parameters in Table~\ref{tab:qubit_params}. At this detuning, $\chi_1$ is smaller than the linewidth of the qubit and individual photon-number states cannot be resolved. The qubit frequency shifts by an average value $\langle n\rangle\chi_1$ of the photon population distribution. 

To spectroscopically resolve individual photon states for Fig.~\ref{fig:main_4} in the main text, we decrease the detuning to $\Delta_{\mathrm{qc}} = -40$~MHz by moving $\omega_{q_0}$ to 5.00~GHz, where $\chi_1 = -7.74$~MHz. 
This operating point is no longer dispersive, and the AC Stark shift shows significant non-linearity. Therefore, both single-photon AC Stark shift $\chi_1 = -7.74$ MHz and $n$-photon AC Stark shifts $\chi_n$ are determined numerically and provided in Fig.~\ref{fig:main_4}.

Because the same qubit is both used to implement the pump and as an ancilla for photon detection, care must be taken to avoid false determinations of $n(mT)$ if the qubit state becomes depolarized and spectroscopy is inconclusive. 
For each spectroscopy trace, we use a Gaussian fit to extract the frequency, width, and depth of the spectroscopy feature corresponding to the qubit transition. We apply automatic selection criteria to reject spectroscopy traces with small fitted amplitude or large background offset, indicative of poor qubit contrast, as well as fits with excessively broadened linewidth, typically indicative of a failed fit. For each pump configuration, these criteria are applied sequentially beginning at cycle $0$, and the analysis is terminated once any criterion is violated. The photon number at each cycle is extracted from the fitted qubit frequency shift.

We define $n_{\rm final}$ of the topological pump as the photon number at the last cycle satisfying all fit criteria, conditioned on systematic pumping occurring at all previous cycles. This last condition is required to reject configurations such as that shown in Fig.~\ref{fig:main_2}(f), where the cavity photon number fluctuates wildly but the qubit still provides sufficient contrast to determine $n(mT)$. For systematic pumping, the expected transfer rate is approximately one photon per cycle. Therefore, for determination of $n_{\rm{final}}$, we reject fits for which the fitted qubit frequency changes by more than $2\chi_1$ between adjacent pump cycles. 
Additionally, we reject fits which indicate a spectroscopy dip, rather than a spectroscopy peak, as these correspond either to failed fits or to cases where the qubit has transitioned to $\ket{e}$, putting the system in a depumping configuration.
If no criteria are violated, we define $n_{\rm{final}}$ as the photon number found after the last cycle.
The orange lines shown in Fig.~\ref{fig:main_2}(b-f) and Fig.~\ref{fig:detuning_sweep_sup}(b-f), indicate the cycles at which the analysis is terminated according to these criteria.

To determine $P$ systematically in both the topological and trivial regimes, we use the same rejection criteria as for $n_{\rm{final}}$, with the exception of allowing fits with inverted contrast in order to capture cases b.1 and b.2 of the topologically trivial regime in which the qubit state flips after every applied drive cycle (see Sec.~\ref{sec:topological_classification}). 
The orange lines in Fig.~\ref{fig:b_sweep_sup}(b-f) indicate the cycle at which the analysis is terminated according to these (slightly reduced) criteria.

For the datasets whose primary objective are to determine $n_{\mathrm{final}}$ for a given configuration rather than the photon-transfer rate per cycle, fits with qubit state ending in $\ket{e}$ are also rejected, since systematic pumping ceases in these scenarios. The orange lines shown in Fig.~\ref{fig:main_2}(b-f) and Fig.~\ref{fig:detuning_sweep_sup}(b-f), indicate the cycles at which the analysis is terminated with the additional criterion.

\end{document}